\documentclass{aa}
\usepackage{graphicx,amsmath,txfonts}

\begin{document}
\newcommand{\mearth}{M_\oplus}
\title{Making giant planet cores: convergent migration and growth of planetary embryos in non-isothermal discs. \\}
\author{A. Pierens \inst{1,2},   C. Cossou  \inst{1,2},  S.N. Raymond \inst{1,2}}
\institute{ Universit\'e de Bordeaux, Observatoire Aquitain des Sciences de l'Univers,
    BP89 33271 Floirac Cedex, France \label{inst1} \and
   Laboratoire d'Astrophysique de Bordeaux,
    BP89 33271 Floirac Cedex, France \label{inst2} \\
     \email{arnaud.pierens@obs.u-bordeaux1.fr}}
\abstract
{
Rapid gas accretion onto gas giants requires the prior formation of $\sim 10\;\mearth$ cores, and this presents a continuing challenge to planet formation models.  Recent studies of oligarchic growth indicate that in the region around 5 AU growth stalls at $\sim 2\; \mearth$. Earth-mass bodies are expected to undergo Type I migration directed either inward or outward depending on the thermodynamical state of the protoplanetary disc.  Zones of convergent migration exist where the Type I torque cancels out. These ``convergence zones'' may represent ideal sites for the growth of giant planet cores by giant impacts between Earth-mass embryos.
}
{
We study the evolution of multiple protoplanets of a few Earth masses embedded in a non-isothermal protoplanetary disc.  The protoplanets are located in the vicinity of a convergence zone located at the transition between two different opacity regimes. Inside the convergence zone, Type I migration is directed outward and outside the zone migration is directed inward.
}
{We used a grid-based  hydrodynamical code  that includes radiative effects. We performed simulations varying the initial number of embryos and tested the effect of including stochastic forces to mimic the effects resulting from disc turbulence. We also performed N-body runs calibrated on hydrodynamical calculations to follow the evolution on Myr timescales.
}
{
For a small number of initial embryos (N = 5-7) and in the absence of stochastic forcing, the population of protoplanets migrates convergently toward the zero-torque radius and forms a stable resonant chain that protects embryos from close encounters. In systems with a larger initial  number of embryos, or in which stochastic forces were included, these resonant configurations are disrupted.  This in turn leads to the growth of larger cores via a phase of giant impacts between protoplanets, after which the system settles to a new stable resonant configuration. Giant planets cores with masses $\ge 10\; \mearth$ formed in about half of the simulations with initial protoplanet masses of $m_p = 3\; \mearth$ but in only 15\% of simulations with $m_p = 1\; \mearth$, even with the same total solid mass.
}
{
If $2-3 \mearth$ protoplanets can form in less than $\sim 1$~Myr, convergent migration and giant collisions can grow giant planet cores at Type I migration convergence zones.  This process can happen fast enough to allow for a subsequent phase of rapid gas accretion during the disc's lifetime.
}

\authorrunning{A. Pierens et al.}
\titlerunning{Convergent migration and growth of planetary embryos in non-isothermal discs}
\maketitle
\section{Introduction}
\label{sec:intro}
The standard scenario for the formation of planets in protoplanetary discs generally involves the following steps: i) coagulation and settling of dust 
in the disc midplane, followed by growth of km-sized planetesimals; ii) runaway growth of planetesimals 
(Greenberg et al. 1978; Wetherill \& Stewart 1989) into $\sim 10^{-5}\;M_\oplus$ embryos; iii) oligarchic growth of these 
embryos (Kokubo \& Ida 1998, 2000; Leinhardt \& Richardson 2005) into planetary cores. Planetary cores forming oligarchically 
beyond the snow-line are expected to have masses $\sim 10\;M_\oplus$ (Thommes et al. 2003) and consequently are able to 
accrete a gaseous envelope to become a giant planet (Pollack et al. 1996) within the lifetime of protoplanetary discs. This however requires a relatively massive protoplanetary 
disc,  equivalent to $10$ times the mass of the minimum-mass solar nebula (hereafter MMSN; Hayashi 1981). Moreover, recent N-body simulations including 
the effects of fragmentation (Levison et al. 2010) indicate only a modest further growth of embryos once these have reached a 
mass of $\sim 2\;M_\oplus$ . This occurs because as the embryos grow, they tend to scatter planetesimals outside 
of their feeding zone rather than accreting them. The 
action of gas drag 
then makes the orbits of these planetesimals quasi-circular, which prevents close encounters with embryos. These results emphasize the 
difficulty of forming giant planet cores within a few Myr in the context of the oligarchic growth scenario.\\
Recently, an alternative model for forming giant planet cores  has been proposed by Lambrechts \& Johansen (2012) and in which 
embryos grow by accretion of cm-sized pebbles.  Compared with the classical scenario, the growth timescale to reach a
critical core mass of $10\;M_\oplus$ is typically reduced by a factor of $30-1000$ at $5$ AU in this model and naturally 
accounts for the preferential prograde spin of large asteroids (Johansen \& Lacerda 2010). Using 
hydrodynamical simulations, Morbidelli \& Nesvorny (2012) 
recently examined this process in more details and found that for a MMSN model, the mass doubling 
time of a $1\;M_\oplus$ embryo accreting $20$-cm pebbles at $1$ AU is only $\sim 5500$ yr. They confirmed 
that the model of Lambrechts \& Johansen (2012) is promising for forming embryos of a  few Earth masses.\\
 Subsequent giant impacts 
between embryos may produce giant planet cores. Giant impacts have been invoked during the late stages of terrestrial planet formation (Wetherill 1985) and may explain the origin of Earth's Moon (e.g., Canup \& Asphaug 2001).  The fact that Uranus' equatorial satellites are on prograde orbits despite the planet's retrograde rotation can be explained by multiple giant impacts of roughly Earth-mass embryos during Uranus' accretion (Morbidelli et al. 2012). Although Neptune's rotation is prograde, its modest obliquity of $28.5\degr$ also appears to require a giant impact (Morbidelli et al. 2012). Finally, it is possible that giant impacts can stimulate runaway gas accretion (Broeg \& Benz 2012). \\
Whether they form by accreting planetesimals or pebbles, embryos must undergo Type I migration 
due to their interaction with the gaseous disc (Ward 1997; Tanaka et al. 2002). The disc torque exerted on a 
low-mass planet and causing Type I migration consists of two components: i) the differential Lindblad torque  due 
to the angular momentum exchange between the planet and the spiral density waves it generates inside the disc, which 
 is generally negative and therefore responsible for inward migration. ii) the corotation torque exerted by the material 
located in the coorbital region of the planet, which scales with both the vortensity (i.e. the ratio between 
the vertical component of the disc vorticity and the disc surface density; Goldreich 
\& Tremaine 1979) and the entropy 
gradients inside the disc (Baruteau \& Masset 2008; Paardekooper \& Papaloizou 2008). In particular, positive surface 
density gradients or negative entropy gradients give rise to a positive  corotation torque which may eventually 
counteract the effect of the differential Lindblad torque and subsequently lead to outward migration. This arises  
provided that an amount of viscous/thermal diffusion is present inside the disc so that the 
corotation torque remains unsaturated, and that diffusion processes operate in such a way that the 
amplitude of the corotation torque is close to its fully unsaturated value. For non-isothermal, viscously 
heated protoplanetary discs,  the torque experienced by a protoplanet is typically positive 
in the inner, optically thicks regions whereas the outer, optically thin regions  give rise to a negative 
torque (Kretke \& Lin 2012; Bitsch et al. 2013). In that case, one can expect protoplanetary discs to present locations 
where the Type I torque cancels and where protoplanets may converge. These are  referred to as zero-migration lines 
or convergences zones and are generally considered as ideal sites for the growth of planetary embryos (Lyra et al. 2010; 
Hasegawa \& Pudritz 2011). \\
A significant body of work has recently investigated the role of these zero-migration lines on the formation of giant 
planet cores.  Sandor et al. (2011) studied this process in isothermal discs where 
zero-torque radii are located at dead-zone boundaries and found that $10\;M_\oplus$ bodies can be formed in 
less than $\sim 5\times 10^5$ yr through collisions of smaller embryos. The case of non-isothermal discs was 
investigated by Hellary \& Nelson (2012) who performed N-body simulations of planetary growth 
in radiatively-inefficient protoplanetary discs. They showed that in non-isothermal discs, the convergent migration induced by corotation torques 
can indeed enhance the growth rate of planetary embryos. Similar results were obtained by Horn et al. (2012) who 
confirmed that giant planet cores can form at convergence zones from sub-Earth mass embryos in $2-3$ Myr.\\
In this paper, we present the results of hydrodynamical simulations of the interaction of multiple protoplanets 
in non-isothermal disc models. Our simulations begin with $3\;M_\oplus$ bodies with positions initially 
distributed around an opacity transition located just inside the snow-line. This opacity transition corresponds to a 
zero-torque radius for planets of masses $1-10\;M_\oplus$, and   
inside (resp. outside) which Type I migration proceeds outward (resp. inward). The main aim of this 
work are: i) to determine the typical evolution outcome of a swarm of mutiple Earth-mass protoplanets 
which convergently migrate toward a zero-migration line ii) to investigate 
whether giant planet cores can be formed at convergence zones from giant impacts between bodies 
of a few Earth masses. With respect to previous studies based on N-body simulations and which employ prescribed forces 
for migration, 
hydrodynamical simulations allow a self-consistent treatment of the interactions between the embryos 
and the gas disc. We performed simulations varying the initial number of embryos and tested the impact 
of including stochastic forces on the planets to mimic the effects resulting from disc turbulence. For laminar 
runs involving a modest initial number of embryos ($N\le 5-7$), we find that the system enters in a long resonant chain which remains 
stable for $\sim 10^4$ yr whereas growth of embryos through collisions occurs when the initial number of objects 
is increased. As expected, including a small level or turbulence tends to break resonant configurations, which 
consequently enhances close encounters between embryos and promotes planetary growth. \\
In order to  study the dynamical evolution on Myr timescales, we  also present in this paper the results of N-body  
runs calibrated on our hydrodynamical simulations. These N-body runs confirm the results of hydrodynamical simulations 
and show that systems which are formed at convergence zones generally reach a quasi-stationary state with each body 
in resonance with its neighbours and evolving on a non-migrating orbit. In $\sim 43\%$ of the  runs 
in which the initial mass of embryos is $\sim 3\;M_\oplus$, protoplanets of masses $\gtrsim 10\;M_\oplus$ were produced, suggesting 
thereby that giant planet cores can be formed at convergence zones through collisions between bodies of a 
few Earth masses. \\
This paper is organized as follows. In Sect. 2, we present the hydrodynamical model. In Sect. 3, we describe how our 
N-body runs are calibrated from hydrodynamical simulations. The results of hydrodynamical simulations are presented 
in Sect. 4. In Sect. 5, we discuss the results of the N-body runs. Finally, we draw our 
conclusions in Sect. 6. 

\section{The hydrodynamical model}
\subsection{Numerical method}
\label{sec:num}
 Simulations were performed with the  GENESIS numerical code (De Val-Borro et al. 2006) which solves the 
 equations for the disc on a polar grid. 
This code employs an advection scheme based on the monotonic transport algorithm 
(Van Leer 1977) and includes the FARGO algorithm (Masset 2000) to avoid time step limitation 
due to the Keplerian orbital velocity at the inner edge of the grid. The energy equation 
that is implemented in the code reads:
\begin{equation}
\frac{\partial e}{\partial t}+\nabla \cdot (e{\bf v})=-(\gamma_{ad}-1)(\nabla\cdot{\bf v})e+Q^+_{\text{visc}}-Q^-_{\text{rad}}
\label{eq:energy}
\end{equation}
where $\bf v$ is the gas velocity, $e$ the thermal energy density and $\gamma_{ad}$ the adiabatic index which is 
set to $\gamma_{ad}=1.4$. Since we expect 
the effects resulting from stellar irradiation to be negligible in the inner parts of 
protoplanetary discs (Bitsch et al. 2013), we include only the contribution from viscous heating 
 in the expression for the heating term $Q^+_{\text{visc}}$. In the previous equation,  
$Q^-_{\text{rad}}$ is the radiative cooling term which is given by:
\begin{equation}
Q^-_{\text{rad}}=2\sigma_B T_{\text{eff}}^4
\label{eq:rcooling}
\end{equation}
where $\sigma_B$ is the Stephan-Boltzmann constant and $T_{\text{eff}}$ is the effective temperature which 
is related to the central temperature $T$ by:
\begin{equation}
T^4=\tau_{\text{eff}}T_{\text{eff}}^4
\end{equation}
where $\tau_{\text{eff}}$ is the effective optical depth given by (Hubeny 1990):
\begin{equation}
\tau_{\text{eff}}=\frac{3\tau}{8}+\frac{\sqrt{3}}{4}+\frac{1}{4\tau}
\label{eq:tau}
\end{equation}
In the previous equation $\tau=\kappa\Sigma/2$ is the optical depth, where $\Sigma$ is the disc surface density 
and $\kappa$ the Rosseland mean opacity which was taken from Bell \& Lin (1994). \\

We employ $N_R=864$ radial grid cells uniformly distributed between $R_{in}=0.6$ and 
$R_{out}=3.2$ and $N_\phi=1800$ azimuthal grid cells. For a $3\;M_\oplus$ planet, this corresponds to the 
half-width of the horseshoe region being resolved by $\sim 5$ grid cells.\\
 We adopt computational units such that the mass 
of the central star is $M_\star=1\;M_\odot$, the gravitational constant $G=1$ and the radius $R_0=1$ in 
the computational domain corresponds to $1$ AU. We use closed boundary conditions at both 
the inner and outer edges of the computational domain and employ wave-killing zones 
for $R<0.75$ and $R>2.7$ to avoid wave reflections at the disc edges.\\

Evolution of planet orbits is computed using a fifth-order Runge-Kutta integrator (Press et al. 1992). 
Close encounter  between the planets $i$ and $j$ is assumed to occur whenever their mutual 
distance is less than $(3m_i/4\pi\rho_s)^{1/3}+ (3m_j/4\pi\rho_s)^{1/3}$, where $m_{i,j}$ is the mass 
of the planets  and $\rho_s$ the mass 
density which is set to $\rho=1\;g.cm^{-3}$. In order to guarantee that two planets do not pass trough each other undetected, 
we set the  
time step size to $\Delta t=min(\Delta t_h, \Delta t_n)$ where $\Delta t_h$ is the hydrodynamical time step and $\Delta t_n$ 
is a N-body timestep given by (Beaug\'e \& Aarseth 1990):
\begin{equation}
\Delta t_n=0.25\min_{i,j}\left|\frac{r_{ij}^2}{\bf{r}_{ij}\cdot\bf{v}_{ij}}\right|
\end{equation}
In the previous expression, ${\bf r}_{ij}=\bf{R_i}-{R_j}$ is the distance between planets $i$ and  $j$ and 
${\bf v}_{ij}={\bf v}_i-{\bf v}_j$ is their relative velocity.\\

Although a 2D disc model is adopted,  we allow planets to evolve in the direction perpendicular to the disc 
midplane as well. With respect to coplanar orbits, this will reduce the collision rate between planets, 
increasing thereby the time during which planets can strongly interact. However, because of the 2D disc model used 
here, bending waves cannot be launched in the disc, and so there is no disc induced damping of inclination as it 
would be in a more realistic 3D disc model. To model the inclination damping due to the interaction with the disc we follow 
Pierens \& Nelson (2008) and mimic the effect of bending waves by applying to each planet with mass $m_p$ a vertical force $F_z$ given 
by:
\begin{equation}
F_z=\beta\frac{m_p\Sigma_p \Omega}{c_s^4}(2A_z^cv_z+A_z^sz\Omega)
\end{equation}
where $c_s$ is the sound speed and where $\Omega$ and $\Sigma_p$ are respectively the Keplerian angular velocity and the disc 
surface density at the position of the planet. $A_z^c$ and $A_z^s$ are dimensionless coefficients which are set to  
$A_z^c=-1.088$ and $A_z^s=-0.871$ (Tanaka \& Ward 2004), and $\beta$ is a free parameter which is chosen such that the 
inclination damping timescale $t_i$ obtained in the simulations is approximately equal to the eccentricity damping 
timescale $t_e$. Test simulations show that choosing $\beta=0.33$ give similar values for $t_i$ and $t_e$.\\

\subsection{Stochastic forces}

 The origin of turbulence is believed to be related to the magneto-rotational instability (MRI, 
Balbus \& Hawley 1991). 
Here, turbulence effects are  modelled as stochastic forcing using the turbulence model of 
Laughlin et al. (2004) and further modified by Baruteau \& Lin (2010). This model employs a turbulent 
potential corresponding to the superposition of $50$ wave-like modes and given by:

\begin{equation}
\Phi_{\text{turb}}(R,\phi,t)=\gamma R^2 \Omega^2\sum_{k=1}^{50}\Lambda_k(R,\phi,t),
\label{eq:phi}
\end{equation}

with

\begin{equation}
\Lambda_k=\xi_k e^{-\frac{(R-R_k)^2}{\sigma_k^2}}
\cos(m_k\phi-\phi_k-\Omega_k\tilde{t_k})
\sin(\pi \tilde{t_k}/\Delta t_k).
\label{eq:lambda}
\end{equation}

In Eq. \ref{eq:lambda}, $\xi_k$ is a dimensionless constant  parameter randomly 
sampled with a Gaussian distribution of unit width. 
$R_k$ and $\phi_k$ are, respectively, the radial and
azimuthal initial coordinates of the mode with wavenumber $m_k$,
 $\sigma_k=\pi R_k /4m_k$ is the radial extent of that mode, and
$\Omega_k$ denotes the Keplerian angular velocity at $R=R_k$.  
 Both $R_k$ and $\phi_k$ are randomly sampled  with a uniform 
distribution, whereas $m_k$ is randomly sampled with a logarithmic
distribution between $m_k=1$ and $m_k=150$.
Each mode of wavenumber $m_k$ starts at time $t=t_{0,k}$ and 
 terminates when $\tilde{t_k}=t-t_{0,k} > \Delta t_k $,
where $\Delta t_k=0.2\pi R_k /m_k c_s$ denotes the lifetime of mode with wavenumber $m_k$. 
Such a value for $\Delta t_k$ yields an autocorrelation time-scale $\tau_c\sim 0.5 T_{orb}$,
where $T_{orb}$ is the orbital period at $R=1$ (Baruteau \& Lin 2010).\\
Following Ogihara et al. (2007),
 we set $\Lambda_k=0$ if $m_k > 6$  to save computing time.
 As noticed by Baruteau \& Lin (2010), such an assumption is supported by the fact that
a turbulent mode with wavenumber $m$ has an amplitude decreasing as $\exp(-m^2)$ and a lifetime
$\propto 1/m$, so that the contribution to the turbulent potential of a high wavenumber turbulent mode
 is relatively weak.
In Eq. \ref{eq:phi}, $\gamma$ denotes the value of the turbulent
forcing parameter and is related to the value of the $\alpha$ viscous stress parameter (Shakura \& Sunyaev 1973) 
by the relation 
(Baruteau \& Lin 2010):
\begin{equation}
\gamma \sim 8.5 \times 10^{-2} h_p \sqrt{\alpha}
\end{equation}
where $h_p$ is the aspect ratio.  Since it is expected the typical amplitude of the surface density 
perturbations to scale with $\gamma$, the previous expression is consistent with the results of 
Yang et al. (2009) who found that these turbulent density pertubations scale with $\sqrt{\alpha}$. Although this parametrisation 
of turbulence does not capture all relevant physical effects like vortices (Fromang \& Nelson 2006) or 
zonal flows (Lyra et al. 2008; Johansen et al. 2009), Baruteau \& Lin (2010) have shown that applying the turbulent 
potential of Eq. \ref{eq:phi} to the disc generates density perturbations that have similar 
statistical properties to those resulting from 3D MHD simulations.

\begin{figure}
\centering
\includegraphics[width=\columnwidth]{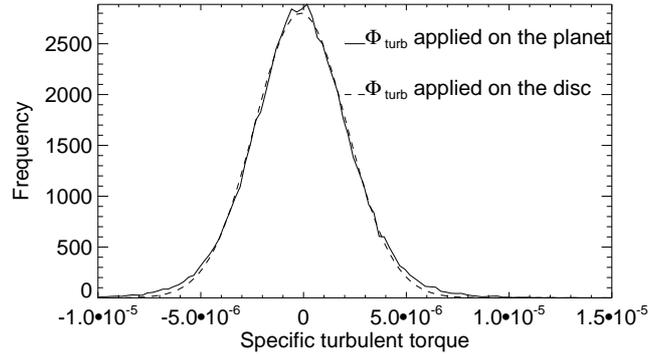}
\caption{Distribution of the specific turbulent torque exerted on a $3\;M_\oplus$ planet obtained from a 
turbulent simulation in which the turbulent potential of Eq. \ref{eq:phi} with $\gamma=10^{-4}$ 
is applied to the disc (solid line), and obtained from a laminar simulation in which the turbulent force of 
Eq. \ref{eq:fturb} with $C=17$ is applied to the planet (dashed line). }
\label{fig:histo}
\end{figure}

In the context of non-isothermal 
disc models, an important effect resulting from these turbulent fluctuations is that the initial temperature  
is progressively alterated by turbulent heating (Pierens et al. 2012).
 In order for the temperature 
profile to remain fixed in the course of simulations, we follow Ogihara et al. (2007) and Horn \& Lyra (2012) 
and rather apply the turbulent potential of Eq. \ref{eq:phi} on the planets. In that case, the 
turbulent force $F_{turb}$ acting on each body is related to $\Phi_{\text{turb}}$ by (Ogihara et al. 2007):
\begin{equation} 
F_{\text{turb}}=-C\frac{64 \Sigma_p R^2}{\pi^2 M_\star}\nabla \Phi_{\text{turb}}
\label{eq:fturb}
\end{equation}
In the previous equation, $C$ is a constant which is set to a value such that the mean deviation of the 
turbulent torque distribution coincides with that obtained in the case where the turbulent 
potential is  applied directly to the disc. In order to  estimate $C$, we have measured the torque experienced 
by a $3\;M_\oplus$ planet for  i) an isothermal 
turbulent simulation in which the turbulent potential of Eq. \ref{eq:phi} with  $\gamma=10^{-4}$ is 
applied to the disc and ii) a series of laminar isothermal simulations in which the turbulent force of Eq. 
\ref{eq:fturb} is applied to the planet, and which differ by the value of $C$. 
In the case with $C=17$, Fig. \ref{fig:histo} shows that the distribution of the specific torque 
acting on the planet is in good agreement with that obtained in the turbulent simulation.

\subsection{Initial conditions}
\label{sec:init}

\begin{figure}
\centering
\includegraphics[width=0.49\columnwidth]{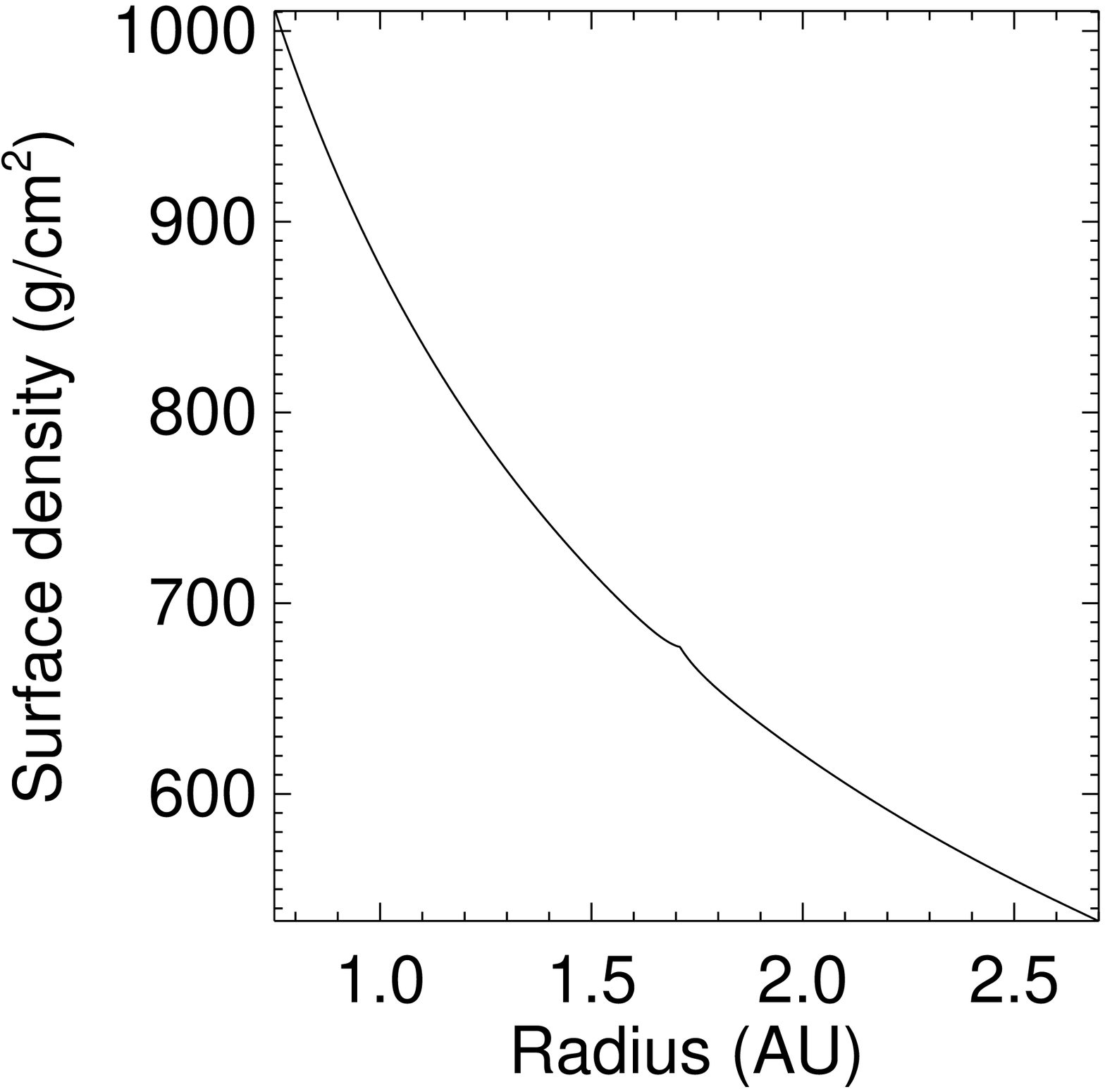}
\includegraphics[width=0.49\columnwidth]{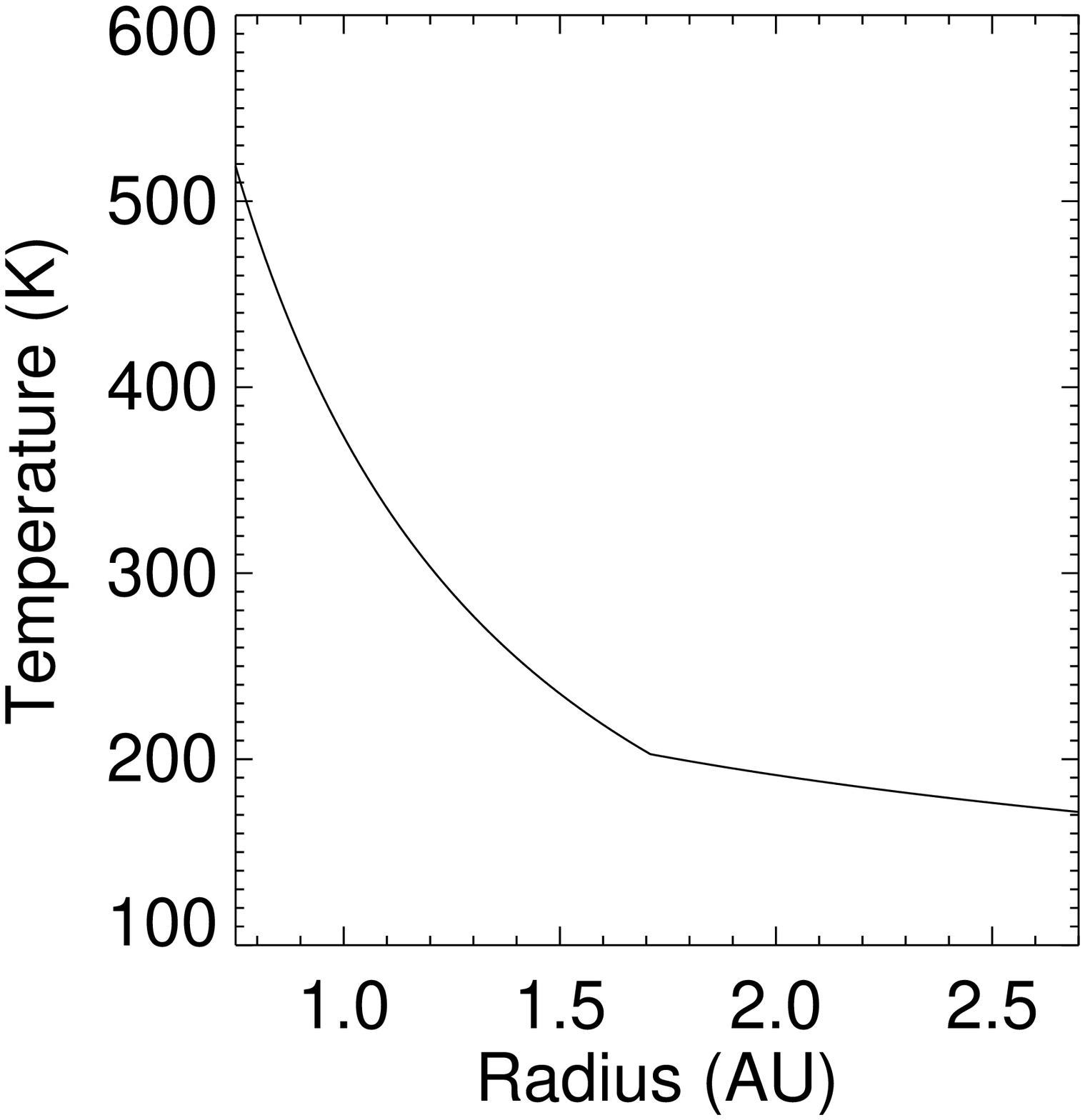}
\caption{Surface density and temperature profiles at steady-state.}
\label{fig:initial}
\end{figure}

The initial disc surface density profile is $\Sigma=\Sigma_p\times(R/R_0)^{-\sigma}$ with 
$\sigma=0.5$  and $\Sigma_p=850$ $g/cm^2$. \\ 
The anomalous viscosity in the disc arising from MHD turbulence  
is modelled 
using a constant kinematic viscosity  $\nu= 10^{14}$ $g/cm^2$, which corresponds to a value for 
the $\alpha$ viscous stress parameter of $\alpha\sim 2\times 10^{-3}$ at $1$ AU. The choice of a 
constant viscosity is justified by the 
fact that for $\sigma=0.5$, there is no viscous evolution of the disc surface density profile so that 
the zero-torque line remains approximately fixed in the course of the simulations. \\
The initial temperature profile 
is $T=T_0(R/R_0)^{-\beta}$ with $\beta=1$ and $T_0$ is the initial temperature at $1$ AU. Under the 
action of the source terms in Eq. \ref{eq:energy}, the temperature profile evolves and reaches an 
equilibrium state once viscous heating equilibrates radiative cooling.
 The surface density and temperature 
profiles at steady-state are plotted in Fig. \ref{fig:initial}. 
The change in the temperature structure at $R\sim 1.7$ AU is related to a change in the opacity regime.
For $R\le 1.7$ AU, the opacity 
is dominated by metal grains and varies as $\kappa \propto T^{1/2}$ whereas for $R\ge 1.7$ AU, melting 
of ice grains causes the opacity to drop with temperature and to vary as $\kappa \propto T^{-7}$ (Bell \& Lin 1994).
By equating the viscous heating and radiative 
cooling terms and assuming an optically thick disc, it is straightforward to show that  $T\propto R^{-8/7}$ inside $R\lessapprox 1.7$ AU 
whereas $T\propto R^{-4/11}$ for $R \gtrapprox 1.7$ AU. This corresponds to an entropy $S=p/\Sigma^{\gamma_{ad}}$, where $p$ is the pressure, 
decreasing as $R^{-0.92}$ inside the opacity transition and as $R^{-0.14}$ for $R \gtrapprox 1.7$ AU. For this disc model, we notice that 
the snow-line is located at $\sim 2.7$ AU, 
which  is consistent with estimations of the location of the snow-line at the epoch of planetesimal formation, although large excursions from this value are expected due to disc evolution (Lecar 2006; Garaud \& Lin 2007).\\

In the hydrodynamical simulations presented below, the initial mass of each protoplanet is assumed to be 
$m_p=3\; M_\oplus$. The motivation for choosing equal-mass embryos is based on the fact that this minimizes the rate of convergent 
migration and therefore the probability of close encounters between embryos when using an isothermal equation of state. 
As we will see in Sect \ref{sec:hydro}, collisions between equal-mass bodies occur in the radiative case but not 
in the locally isothermal limit, which clearly demonstrates the role of the zero-migration line in forming 
more massive objects through collisions. Isothermal simulations with embryos of initially different masses would 
probably lead to close encouters between embryos (Cresswell \& Nelson 2006).\\
 The distribution of semi-major axes $a_i$ is such that about half of initial population of planetary embryos  
is located on each side of the convergence zone, with an 
initial orbital separation between two adjacent bodies $p$ and $p'$ of $4.5\; R_{mH}$,  where $R_{mH}$ is the mutual Hill radius:
\begin{equation}
R_{mH}=\frac{a_p+a_{p'}}{2}\left(\frac{m_p+m_{p'}}{3M_\star}\right)^{1/3}
\end{equation}
where $a_{p,p'}$ denotes the semimajor-axes of planets $p$ and $p'$.
Although the initial separation between bodies is greater than the critical value of $3.46\;R_{mH}$ below 
which rapid instability occurs for two planets on initially circular orbits (Gladman 1993),  it is smaller to 
what is expected from the oligarchic growth scenario (Kokubo \& Ida 1998). The adopted value for the initial 
embryo separation is chosen to make the hydrodynamical simulations computationally tractable, but N-body 
runs performed  with an initial separation of $6\; R_{mH}$ show consistent results in comparison with 
those obtained using the fiducial value of $4.5\; R_{mH}$ (see Sect. \ref{sec:nbody}).\\
Planetary embryos initially evolve on  circular orbits with inclinations randomly sampled according 
to  a Gaussian distribution with mean $\mu_i=0^\circ$ and standard deviation $\sigma_i=1^\circ$.

\section{Calibration of N-body simulations}
\label{sec:calibration}
\subsection{Prescription for Type I migration}
\begin{figure}
\centering
\includegraphics[width=\columnwidth]{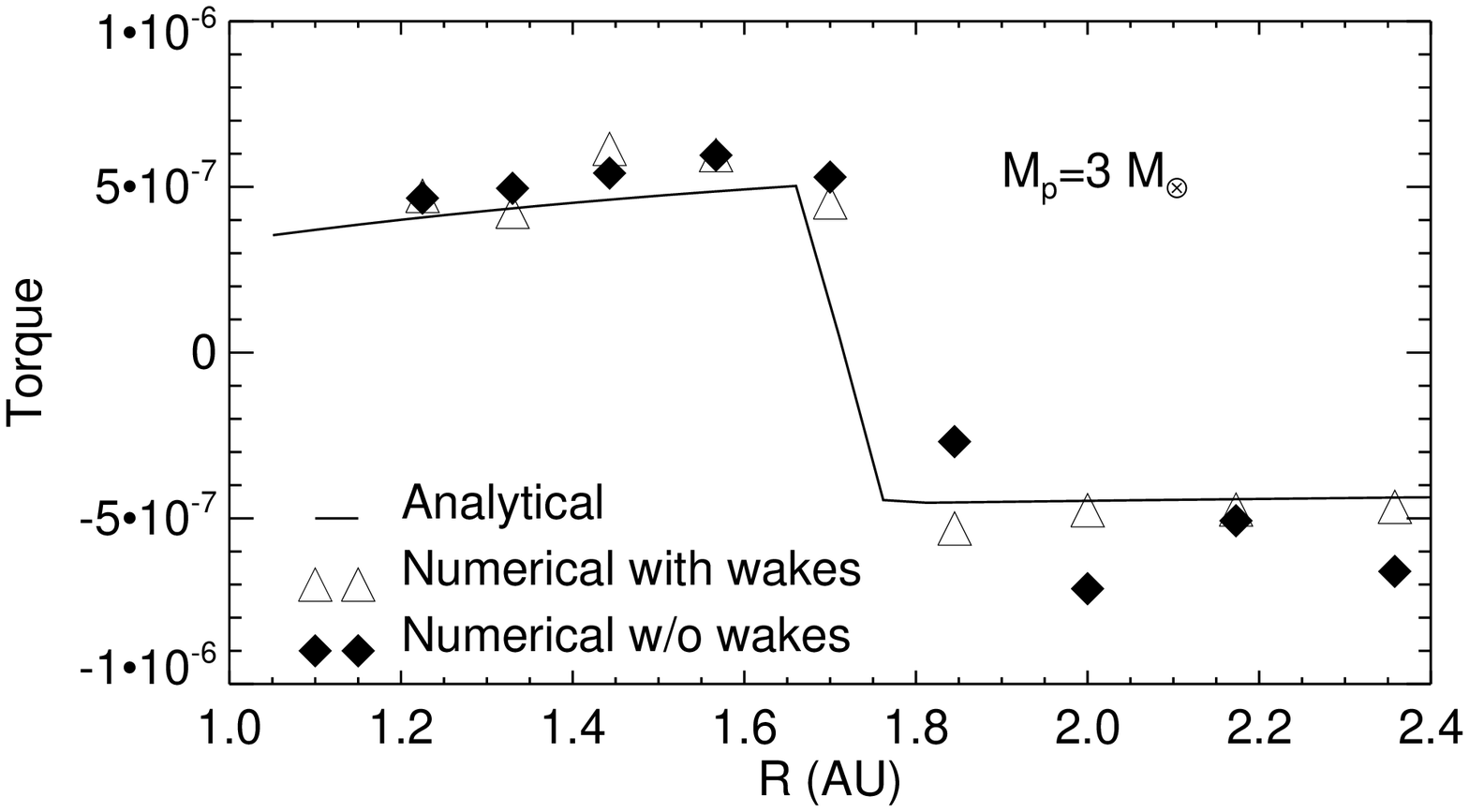}
\includegraphics[width=\columnwidth]{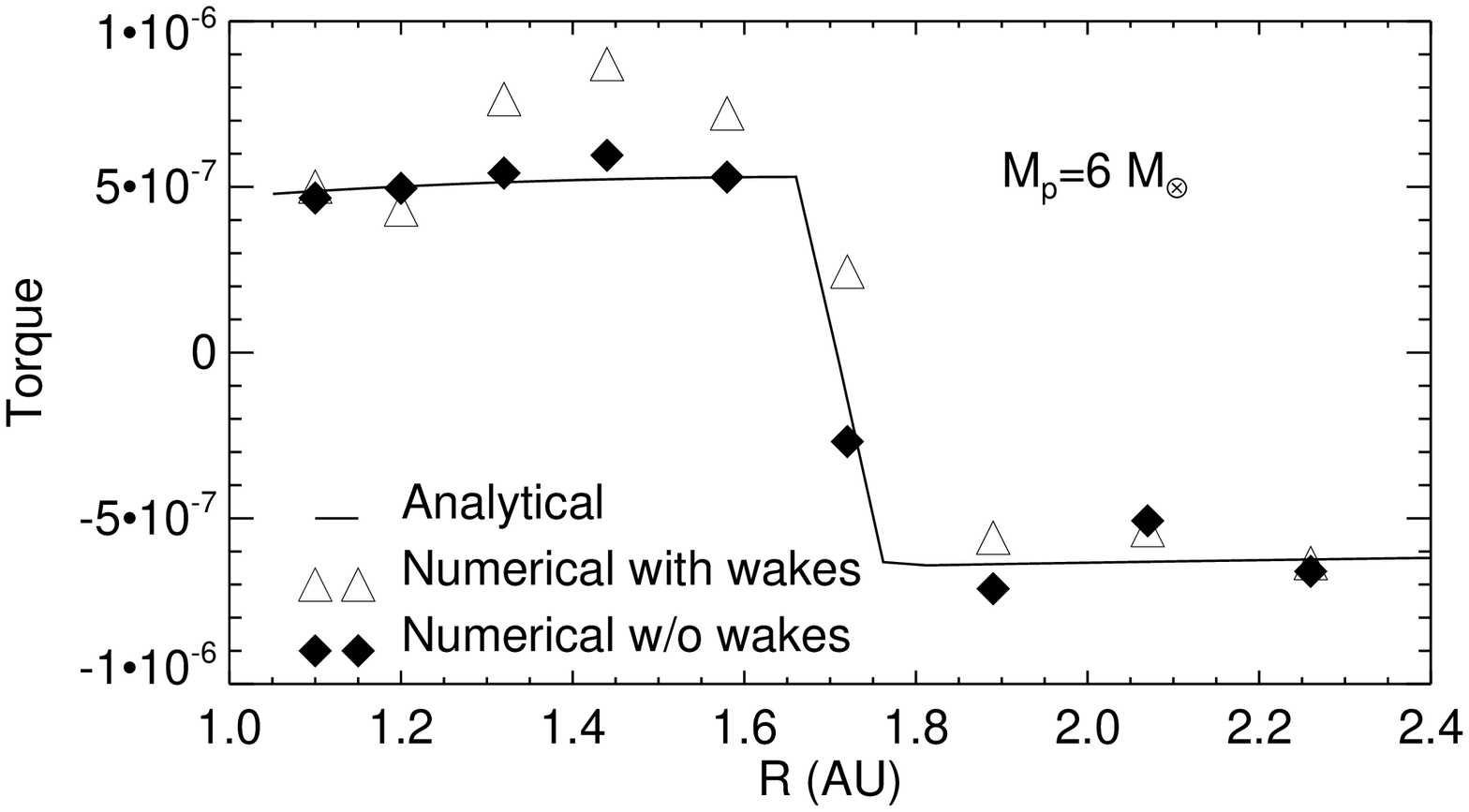}
\caption{{\it Upper panel:} Torque exerted on a $3\; M_\oplus$ planet as a function of its orbital distance. 
The solid line 
corresponds to the analytical prescription of Paardekooper et al. (2012). Triangles and diamonds are the results 
from hydrodynamical simulations with/without the effects due to the waves generated by the other planets included. 
{\it Lower panel:} same but for $m_p=6\;M_\oplus$.}
\label{fig:radtorque}
\end{figure}

The torque exerted on a protoplanet embedded in a non-isothermal disc can be decomposed into two components: the 
differential Lindblad torque which results from angular momentum exchange between the planet and the spiral waves 
it generates inside the disc plus the corotation torque which is due to the torque exerted by the material located 
in the coorbital region of the planet. A linear analysis reveals that the corotation torque consists of a barotropic part 
which scales with the vortensity plus an entropy-related part which scales with the entropy gradient. In the 
absence of any diffusion processes inside the disc, however, vortensity and entropy gradients tend to flatten through 
phase mixing, which causes the two components of the corotation torque to saturate. Consequently, desaturating the 
corotation torque requires that some amount of viscous and thermal diffusions are operating inside the disc. In that 
case, the amplitude of the corotation torque depends on the ratio between the diffusion time-scales and the horseshoe 
libration time-scale. Its optimal value, also referred to as the fully unsaturated corotation torque, is obtained when the diffusion time-scales are approximately equal to 
half the  horseshoe libration time (e.g. Baruteau \& Masset 2013). In the limit where the diffusion time-scales become 
shorter than the U-turn time-scale, the corotation torque decreases and approaches the value predicted by linear 
theory. Therefore, the corotation torque can be considered as a linear combination of the fully unsaturated corotation torque 
and the linear corotation torque, with coefficients depending on the ratio between the diffusion time-scales and the 
horseshoe libration time-scale. Torque formulae as a function of viscosity and thermal diffusivity were proposed by 
Paardekooper et al. (2012).\\

In our N-body runs, Type I migration is modelled by an extra-force $\bf{a_m}$ acting on each body and defined by:
\begin{equation}
{\bf a}_m=-\frac{{\bf v_p}}{t_m}
\end{equation}
In the previous expression, $\bf{v_p}$ is the planet velocity and $t_m=J/2\Gamma$ is the migration timescale, 
where $J$ is the specific planet angular momentum and $\Gamma$ the specific disc torque.  To estimate $\Gamma$, we use the 
analytical  prescription of Paardekooper et al. (2012) but  multiplied by a factor of $0.7$. As we will see shortly, very good agreement with hydrodynamical simulations is obtained in that case. We notice that this is consistent with 
the results of Cresswell \& Nelson (2006) who found that for isothermal disc models, analytical torque 
formulae given in Tanaka et al. (2002) predict migration times that are faster than those observed in the simulations 
by about a factor of three. \\
For a purely active disc, radiative diffusion and viscous time-scales are expected to be equal 
(Bitsch \& Kley 2011) so when computing the saturation parameters in the analytical formulae, we 
 set $\nu=\chi$, where $\chi$ is the thermal diffusion coefficient. In the case where thermal diffusion is only due to 
radiative effects, $\chi$ is given by (e.g. Bitsch \& Kley 2011):
\begin{equation}
\chi=\frac{16\gamma_{ad}(\gamma_{ad}-1)\sigma T^4}{3\kappa \rho^2 H^2 \Omega^2}
\end{equation}
where $H$ is the disc scale height and $\rho=\Sigma/2H$ the gas density.
In Fig. \ref{fig:radtorque}, we compare for the disc model described in Sect. \ref{sec:init} and for  $m_p=3,\; 6\; M_\oplus$ the 
analytical torque of Paardekooper et al. (2012), which we multiplied by a factor of $0.7$, with 
the numerical torque obtained using GENESIS. 
 Clearly, very good agreement is obtained between the analytical
 prediction and the torques derived from numerical simulations.
 Inside the zero-torque radius located at $R\sim 1.7$ AU, the torque 
is positive due to the strong (negative) entropy gradient there ($S\propto R^{-0.92}$, see Sect. \ref{sec:init}) whereas outside the zero-migration line, the entropy gradient is weaker ($S\propto R^{-0.14}$) so that the 
(positive) entropy-related corotation torque is not strong enough to counterbalance the (negative) differential Lindblad torque. We note that 
in Fig. \ref{fig:radtorque}, the zero-migration line at $R\sim 1.7$ AU exists provided that the corotation torque remains unsaturated. This arises when the diffusion 
timescale across the horseshoe region $t_{dif}=x_s^2/\chi$ is shorter than the libration timescale 
$t_{lib}=8\pi a_p/3\Omega_p x_s$ but longer than the U-turn timescale $t_{U-turn}\sim h_p t_{lib}$, where $x_s$ is the half-width 
of the horseshoe region which is given by (Paardekooper et al. 2010):
\begin{equation}
x_s=\frac{1.1a_p}{\gamma_{ad}^{1/4}}\sqrt{\frac{q}{h_p}}
\end{equation}
where $q=m_p/M_\star$ the planet mass ratio. This condition yields an estimation of the planet mass range for which 
the corotation torque remains unsaturated. We find:
\begin{equation}
2.8\left(\frac{\chi}{a_p^2\Omega_p}\right)^{2/3}h_p^{5/3}\lesssim q\lesssim 2.8\left(\frac{\chi}{a_p^2\Omega_p}\right)^{2/3}h_p
\end{equation}
which gives $3.5\times 10^{-6}\lesssim q\lesssim 3\times 10^{-5}$ for our disc model. This implies that convergent 
migration toward the opacity transition is expected for planets with masses in the range 
$1 \lesssim m_p \lesssim 10\;M_\oplus$. Bodies more massive than $\sim 10\;M_\oplus$ or less 
massive than $\sim 1\;M_\oplus$ will  rather experience 
inward migration.\\

We also  examined the issue of whether the torque experienced by a protoplanet can be altered by the close 
proximity of other bodies. Horn \& Lyra (2012) indeed speculated that effects resulting from planet wakes 
may lead to a more rapid saturation of the corotation torque. To achieve this aim, we performed i) one simulation in which 
we measured the torques experienced by N=9 
protoplanets separated by $4.5\;R_{mH}$  and held on a fixed circular orbit, ii) an additional set of calculations with 
$N=1$ protoplanet and which differ in the value for the planet  radial position. From Fig. \ref{fig:radtorque} 
we see that the torques derived from these two series of runs differ only slightly, which suggests that the wakes generated 
by other low-mass planets have only a marginal effect on the saturation of the corotation torque.\\

\begin{figure}
\centering
\includegraphics[width=\columnwidth]{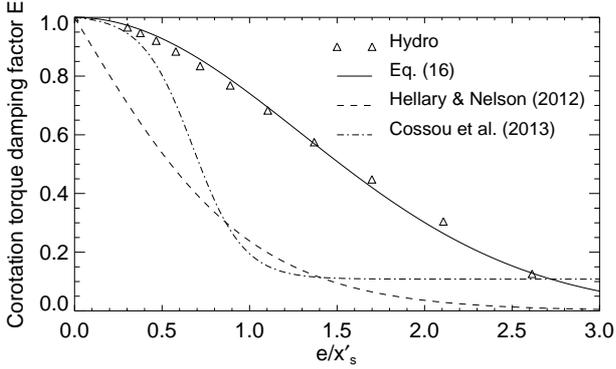}
\caption{Corotation torque damping factor as a function of the ratio between the eccentricity and the dimensionless half-width of the horseshoe region for 
$m_p=3\;M_\oplus$, and deduced from hydrodynamical simulations (triangles). The solid line corresponds to the 
function that best fits the simulation results. The dashed line is the prescription of Hellary \& Nelson (2012) and 
the dash-dotted line corresponds to the prescription of Cossou et al. (2013).}
\label{fig:damping}
\end{figure}

When the planet eccentrity reaches a value such that its radial excursion bebomes larger than the half-width of the 
horseshoe region, we expect the corotation torque to be 
strongly attenuated. To model this effect, we follow Hellary \& Nelson and multiply in the N-body runs  the 
analytical corotation torque by 
a damping factor $E$. In order to estimate how $E$ depends on $e_p$, we have performed a subset of calculations  
with a $m_p=3\;M_\oplus$ protoplanet evolving on a fixed circular orbit with  $e_p$ in the range $0\le e_p\le h_p$. Since
the half-width of the horseshoe region is a fraction of the disc scaleheight, we expect a null corotation torque 
when $e_p \sim h_p $, leaving only the differential Lindblad torque. Given that the differential Lindblad torque 
depends weakly on the eccentricity for $e_p<h_p$, the corotation torque 
can be determined in each simulation by simply substracting the differential Lindblad torque to the total torque.  
We plot the results of these simulations in Fig. \ref{fig:damping}. Superimposed is the function that is found 
to best reproduce
the simulation results and which is given by:
\begin{equation}
E=\exp\left(-\frac{e_p^2}{X_s'^2}\right)
\end{equation}
In the previous equation, $X'_s$ is defined by $X_s'=2x'_s$,  where $x'_s=x_s/a_p$ is the dimensionless half-width of the horseshoe region.
Compared with the  prescriptions of Hellary \& Nelson (2012) and Cossou et al. (2013) which are also plotted in Fig. \ref{fig:damping}, our formula
tends to enhance the corotation torque by a factor of $\sim 3$ for $e_p\sim x'_s$.   

\subsection{Eccentricity and inclination damping in N-body simulations}
\begin{figure}
\centering
\includegraphics[width=\columnwidth]{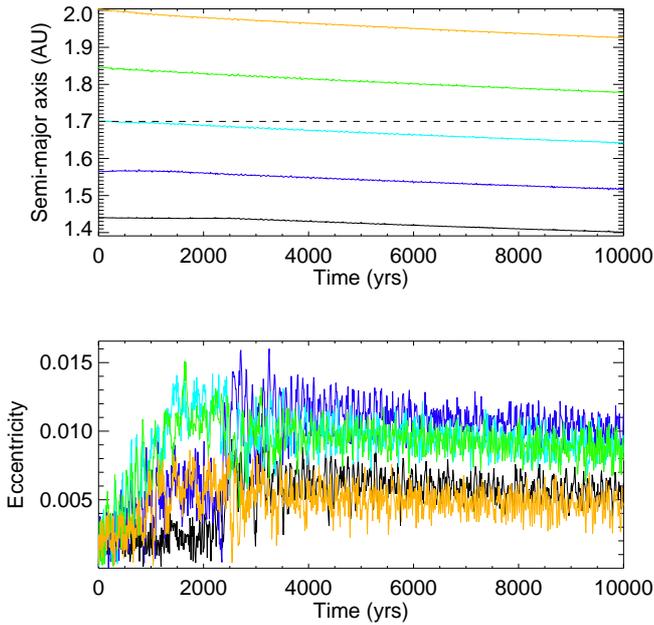}
\caption{Orbital evolution of N=5 embryos with mass $m_p=3\;M_\oplus$ and without stochastic forces included. The dashed line
 shows the location of the zero-migration line.} 
\label{5mp3}
\end{figure}
\begin{figure}
\centering
\includegraphics[width=\columnwidth]{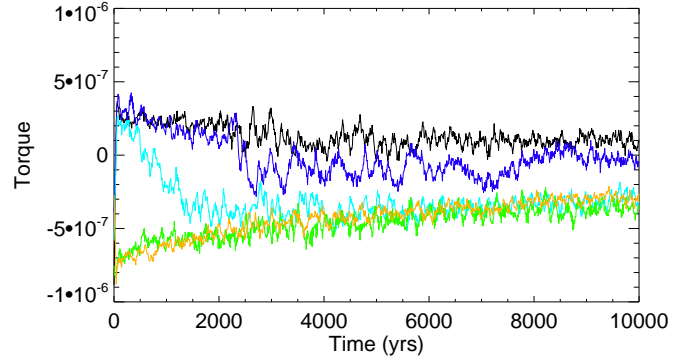}
\caption{Time evolution of the torque experienced by the protoplanets in the case with N=5 and without 
stochastic forces included}
\label{fig:torque_5mp3}
\end{figure}

Eccentricity and inclination damping resulting from the interaction with the disc are modelled by applying the 
following accelerations to each body:
\begin{equation}
{\bf a}_e=-2\frac{({\bf v}_p\cdot{\bf R}_p){\bf R}_p}{R_p^2 t_e}
\end{equation}
and
\begin{equation}
{\bf a}_i=-\frac{({\bf v}_p\cdot \hat{\bf z})}{t_i} \hat{\bf z}
\end{equation}
where $\hat {\bf z}$ is a unit vector in the vertical direction. $t_e$ and $t_i$ are the eccentricity 
and inclination damping timescales for which we use the prescriptions of Cresswell \& Nelson (2008):
\begin{equation}
t_e=\frac{t_{wave}}{0.78}\left[1-0.14\left(\frac{e_p}{h_p}\right)^2+0.06\left(\frac{e_p}{h_p}\right)^3+
0.18\left(\frac{e_p}{h_p}\right)\left(\frac{i_p}{h_p}\right)^2\right]
\end{equation}
and
\begin{equation}
t_i=\frac{t_{wave}}{0.544}\left[1-0.3\left(\frac{i_p}{h_p}\right)^2+0.24\left(\frac{i_p}{h_p}\right)^3+
0.18\left(\frac{e_p}{h_p}\right)^2\left(\frac{i_p}{h_p}\right)\right]
\end{equation}
where:
\begin{equation}
t_{wave}=\frac{M_\star}{m_p}\frac{M_\star}{\Sigma_p a_p^2}h_p^4\Omega_p^{-1}
\end{equation}

\section{Results of hydrodynamical simulations}
\label{sec:hydro}
\subsection{Laminar viscous simulations}
\begin{figure*}
\includegraphics[width=0.49\textwidth]{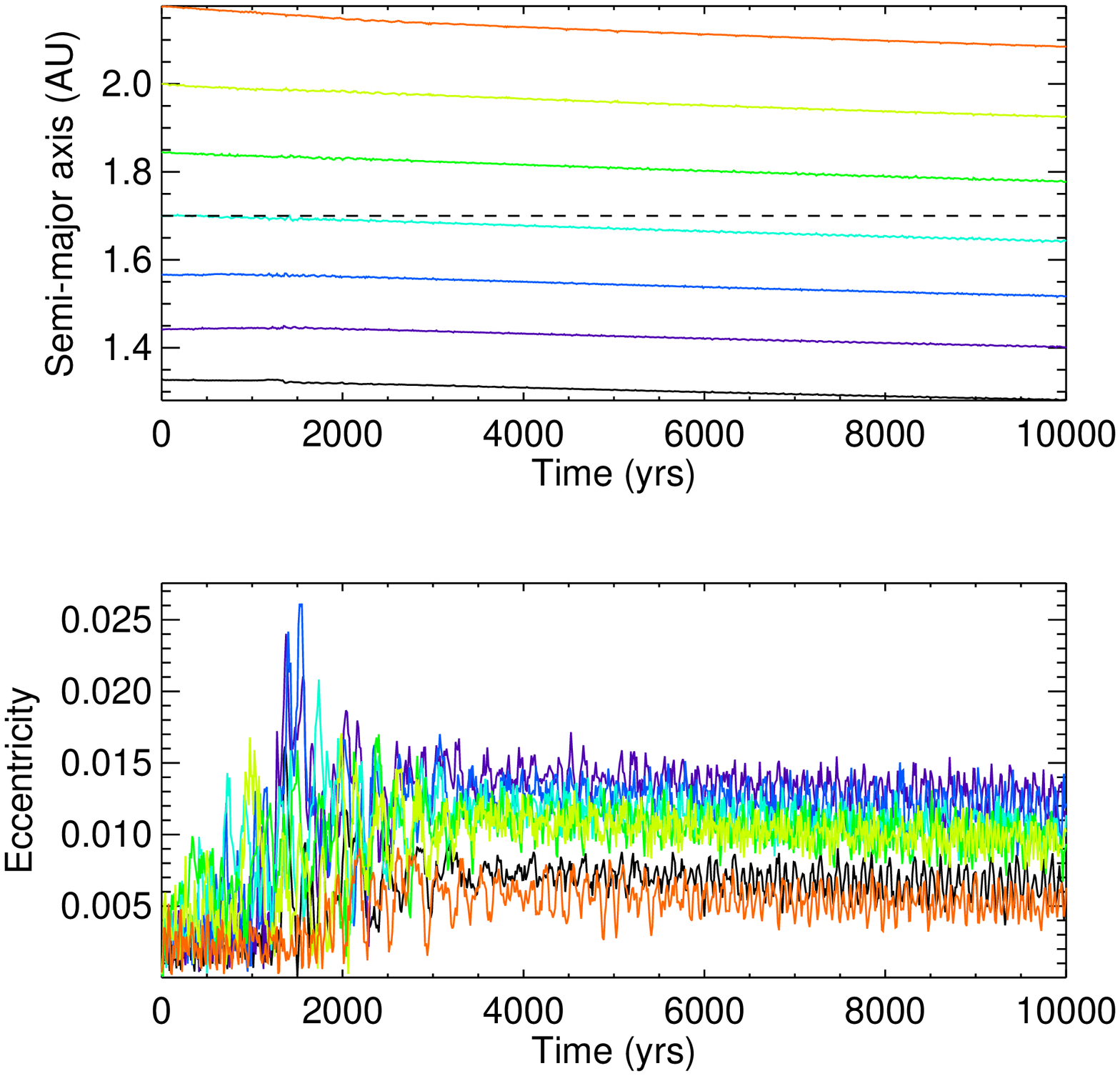}
\includegraphics[width=0.49\textwidth]{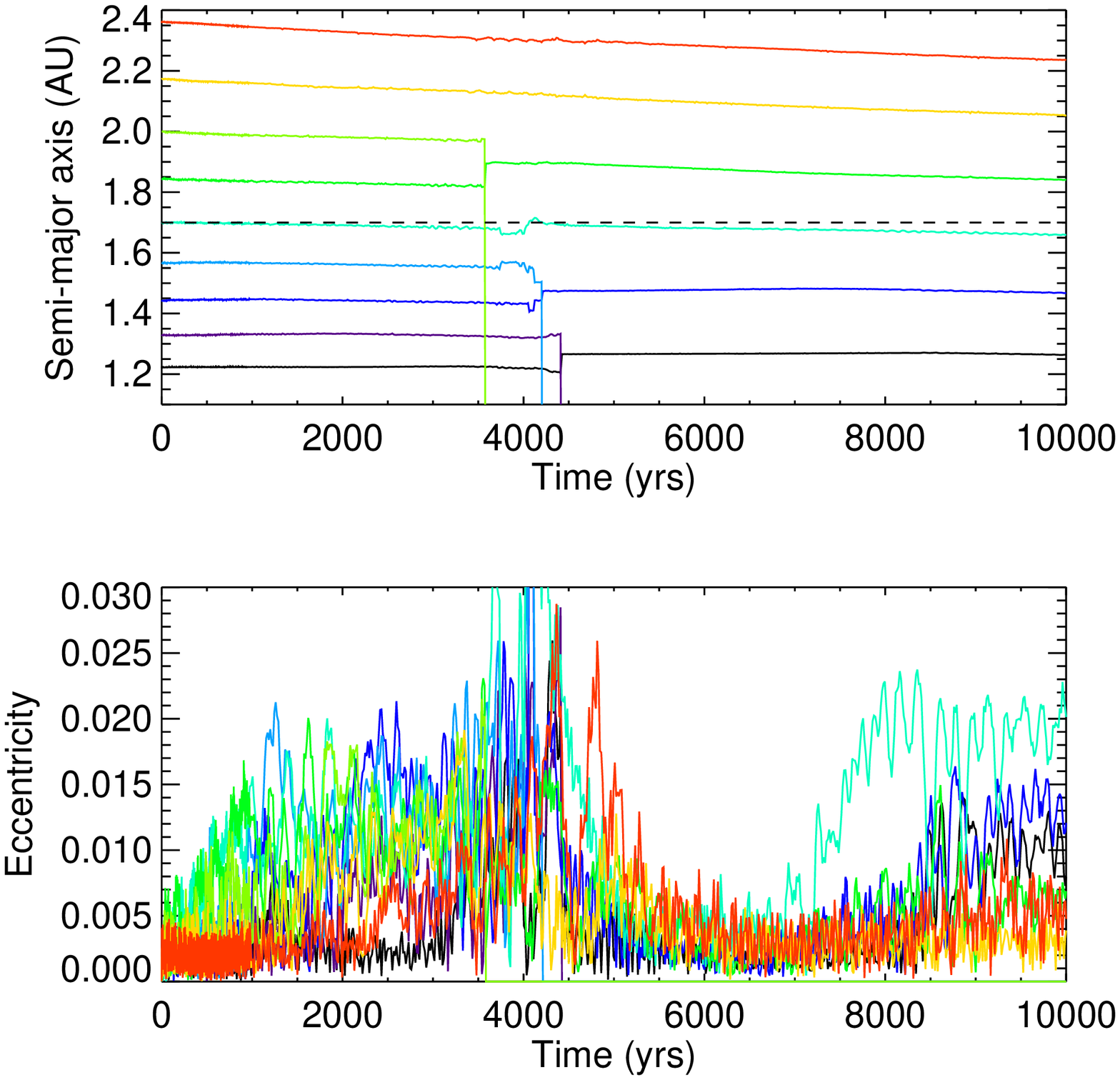}
\caption{{\it Left panel:} Orbital evolution of N=7 embryos with mass $m_p=3\;M_\oplus$ and without stochastic forces included.  
{\it Right panel:} same but with N=9.  The dashed line
 shows the location of the zero-migration line.}
\label{7mp3}
\end{figure*}

The orbital evolution of N=5 embryos with mass $m_p=3\; M_\oplus$ and initially separated by $4.5\; R_{mH}$ is displayed in Fig. \ref{5mp3}. At early times, the two innermost (resp. outermost)  bodies located inside (resp. outside) the opacity transition tend to undergo outward (resp. inward) migration. Regarding the third body (cyan) initially located at $R_p=1.7$, it tends to experience 
only a weak 
positive torque due to its close proximity to the opacity transition. Nevertheless, the strong differential migration between
 this body and the fourth 
one (green) quickly leads to the formation of a 9:8 mean motion resonance (MMR) between these two protoplanets 
at $t\sim 3\times 10^2$ yr. 
As revealed by Fig. \ref{fig:torque_5mp3} which displays the temporal evolution of the torques experienced by each planet, 
eccentricity growth due to resonant trapping makes the torque 
experienced by the third planet become negative, in such a way that the third and fourth planets subsequently migrate inward together 
while maintaining their 
9:8 resonance. This arises because, just after resonant trapping, the value reached by the eccentricity of the third planet 
$e_p\sim 0.015$ is comparable to the 
dimensionless half-width of the planet's horseshoe region which is estimated to be $x'_s\sim 1.1 a_p\sqrt{q/h}\sim 0.02$. 
Consequently, the radial excursion that the planet undergoes is eventually  larger than the horseshoe region, 
causing thereby the (positive) corotation torque to be significantly attenuated (Bitsch \& Kley 2010, 
Cossou et al. 2013). \\
These two bodies then catch up with the outward-migrating second protoplanet (blue) and enter in a 9:8 MMR with it at $t\sim 600 $ yr. Again, eccentricity growth due to resonant capture causes the positive torque exerted on the second body to weaken. Although 
it remains positive, its amplitude is not sufficient to counterbalance the negative torques exerted on the 
third and fourth planets, and this three-planet system consequently suffers a slow, inward resonant migration. This proceeds 
until the fifth body (orange) catches up with the fourth protoplanet (green) and enters a 9:8 resonance with it at $t\sim 1000$ 
yr. At $t\sim 2000$ yr, the outward-migrating innermost planet (black) becomes trapped in a 9:8 MMR with the second 
protoplanet (blue) which, from this time onward, undergoes a marginally positive torque due to the high value reached by 
its eccentricity. However, as can be seen in the lower panel of Fig. \ref{5mp3}, eccentricity 
pumping due to resonant interaction is relatively modest for the innermost  core, with an equilibrium value of 
$e_p\sim 5\times 10^{-3}$. Consequently,  the fraction of the corotation 
torque operating on the innermost planet is large enough for the total torque exerted on this body to remain positive, 
which is confirmed by looking at the evolution of the torque exerted on that planet in Fig. \ref{fig:torque_5mp3}.
This effect, however, is clearly not sufficient to couterbalance the negative torques experienced by the outer bodies 
so that at late times, the five bodies  tend migrate inward in 
lockstep with each neighbouring pair  of planets forming a 9:8 resonance.  As the outer planets pass through the 
zero-torque radius, however, we expect the disc torques exerted on these planets to become positive so that it is likely that 
the swarm will ultimately stop migrating once the net torque acting on the whole system cancels (Cossou et al. 2013). \\

\begin{figure}
\centering
\includegraphics[width=\columnwidth]{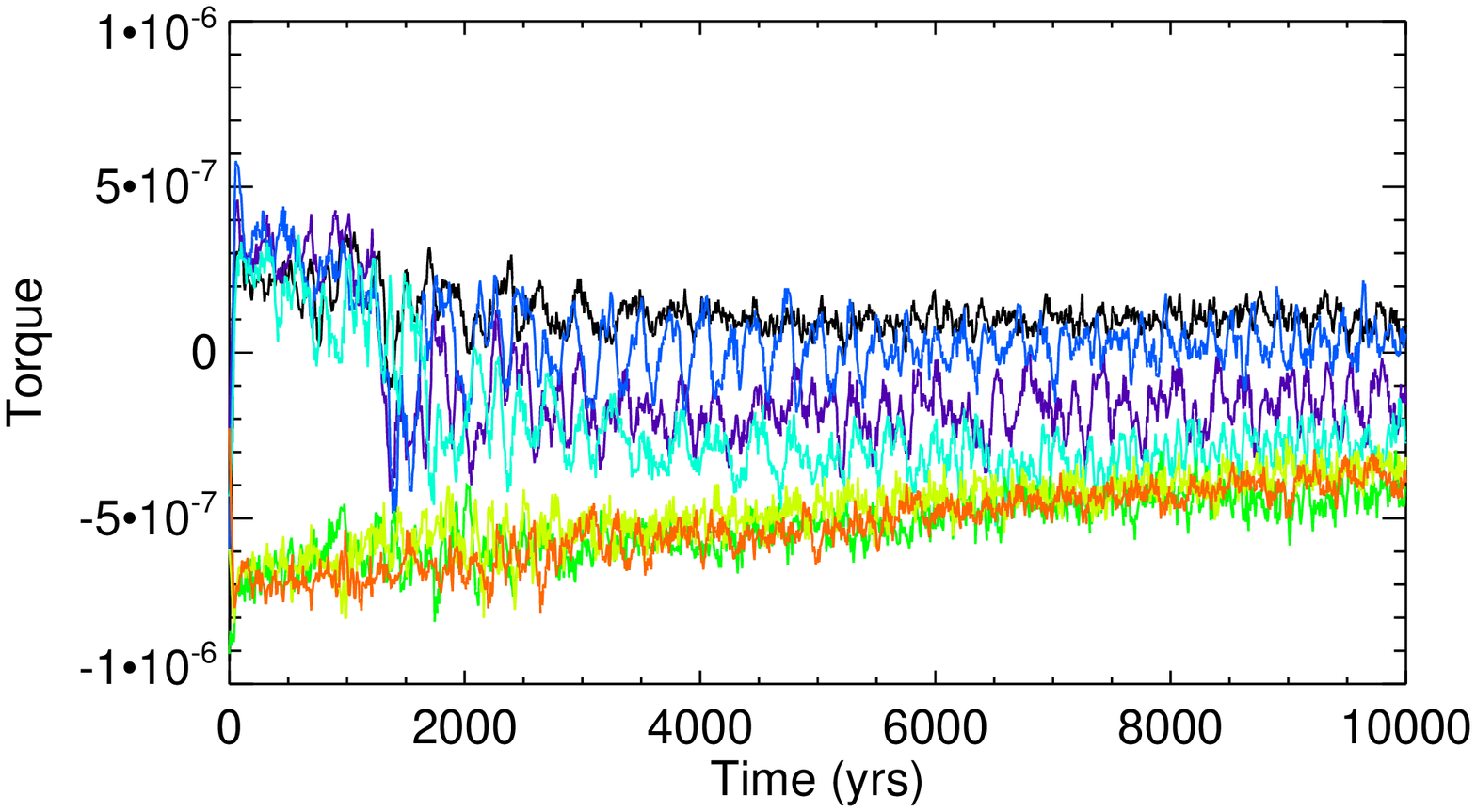}
\includegraphics[width=\columnwidth]{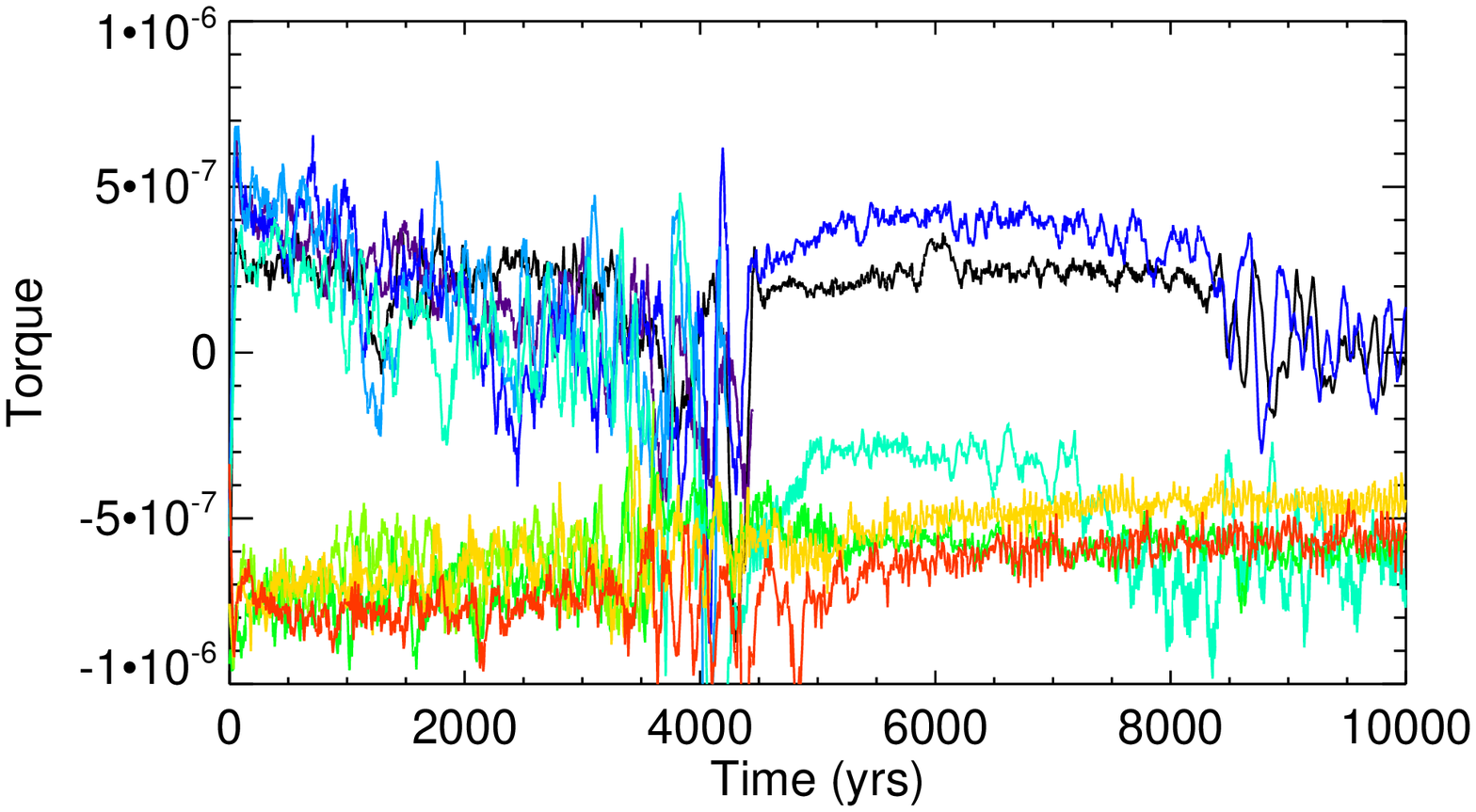}
\caption{{\it Upper panel:}Time evolution of the torque experienced by the protoplanets in the case with N=7 and 
without stochastic forces included. {\it Lower panel:} same but for N=9.}
\label{fig:torque_7mp3}
\end{figure}

In Fig. \ref{7mp3} we present, the orbital evolution of simulations with N=7 (left panel) and N=9 (right panel)
embryos embedded in the same disc model. We remind the reader that the embryos are initially located in such a way that bodies 
of the inner half migrate outward whereas bodies of the 
outer half   migrate inward. Because they are initially located on either side of the convergence line, 
the fourth (cyan) and fifth (green) bodies are the first to become trapped in a 9:8 MMR. Again, eccentricity growth due to resonant trapping 
causes the corotation torque operating on the fourth body to be significantly attenuated so that these two planets tend to 
migrate inward at later times. This is exemplified in the upper panel of Fig. \ref{fig:torque_7mp3} which shows the evolution of the 
torque exerted on each planet as a function of time.  At $t\sim 600$ yr, the third body (blue) catches up with the fourth body 
(cyan) and enters 
a 9:8 MMR with it. A general trend is that inward-migrating bodies are captured in resonance later than 
outward-migrating bodies. This is a direct consequence from the  corotation damping effect, which tends to make the resonant swarm 
migrate inward, strengthening thereby the differential migration rate with the inner, outward-migrating bodies.\\
  Once again, the evolution outcome consists of inward migration of a group of N=7 members 
which are in mutual mean motion resonances, with the resonance being  9:8 except for the two innermost bodies which are 
in 8:7 resonance. This arises because prior to resonant capture of the second body (purple) , the torque exerted on this planet 
is slightly stronger in comparison with that experienced by the innermost one (black), resulting in divergent migration between 
these two cores. Comparing Figs. \ref{fig:torque_5mp3} and \ref{fig:torque_7mp3}, we see that after $\sim 10^4$ 
 yr,  only one planet feels a positive torque in the case where N=5 while two cores are subject to a positive 
torque in the simulation with N=7. This confirms the expectation that the resulting resonant system becomes more compressed as $N$ increases, and consequently more prone to dynamical instability.\\

Indeed, increasing the number of initial embryos to N=9 resulted in a more chaotic behaviour where protoplanets 
suffered close encounters and collisions, as illustrated in the right panel of Fig. \ref{7mp3} which displays the 
planets' orbital evolution for that case. At early times, evolution proceeds similarly to that corresponding to 
N=7, with a system of 9 protoplanets with each body locked in a 
9:8 MMR with its neighbours being formed at $t\sim 3000$ yr. From that time onward, it can be seen in the 
lower panel of Fig. \ref{fig:torque_7mp3}, which displays the temporal evolution of the torque exerted on 
each planet, that the five innermost embryos located inside the convergence line undergo 
a positive torque whereas the others undergo a negative torque, implying a significantly compressed resonant 
system. At $t\sim 3500$ yr, this leads to  a physical collision between the sixth (green) and seventh (light green) embryos, 
forming thereby a $6\; M_\oplus$ planet which subsequently undergoes inward migration since it is located 
outside the convergence line. 
The resulting perturbation then propagates to the inner system and causes two 
additional collisions at later times,  between the third (dark blue) and fourth (blue) planets at $t\sim 4000$ yr and between the two innermost cores (black+purple)  at $t\sim 4100$ yr. As revealed by the lower right panel of Fig. \ref{7mp3}, these two newly 
formed $\sim 6\; M_\oplus$ planets are located inside the convergence line and have relatively modest eccentricity. This, combined with the fact that 
the fully-unsaturated corotation torque scales as $m_p^2$ (or equivalently as $x_s^4$) for low-mass planets, implies that these planets are subject to a stronger positive torque. Not surprisingly, the significant resulting differential migration between the outermost $6\; M_\oplus$ planet (blue) and its 
exterior inward-migrating $3\; M_\oplus$ neighbour (cyan) leads to capture in a 6:5 resonance at $t\sim 7000$ yr. Eccentricity 
growth due to resonant trapping causes the corotation torque operating on the $6\; M_\oplus$ to be partially attenuated, causing 
the two planets migrate inward together at later times while maintaining their 6:5 resonance. This process enables the innermost 
$6\;M_\oplus$ planet (black) to catch up with the outermost $6\;M_\oplus$ body (blue)  and to enter in a 5:4 resonance with it at 
$t\sim 8500$ yr. Protoplanets located outside the convergence line then become trapped in MMRs with the inner ones at 
later times so that the evolution outcome for that case corresponds again to the formation of a stable resonant system where the two inner 
planets are the more massive. Since the two inner planets are trapped in resonance with 
sustained eccentricities, they only feel a weak positive torque despite being located interior to the 
convergence zone. With no strong outward push, the whole system of embryos migrates inward. The high 
computational cost makes it impossible to perform hydrodynamical simulations for more than $\sim 10^4$ yr so that the 
final fate of the system remains uncertain. Altough it can not be excluded that additional collisions will arise 
at later times, a possible issue is that the planets will reach stationary orbits once the net torque acting on 
the resonant system, which consists of the sum of the attenuated corotation torques and unattenuated differential 
Lindlbad torques exerted on each planet, cancels (Cossou et al. 2013). In Sect. \ref{sec:nbody}, we will examine in more 
details the possible long-term evolution outcomes using N-body simulations. \\

\begin{figure}
\centering
\includegraphics[width=\columnwidth]{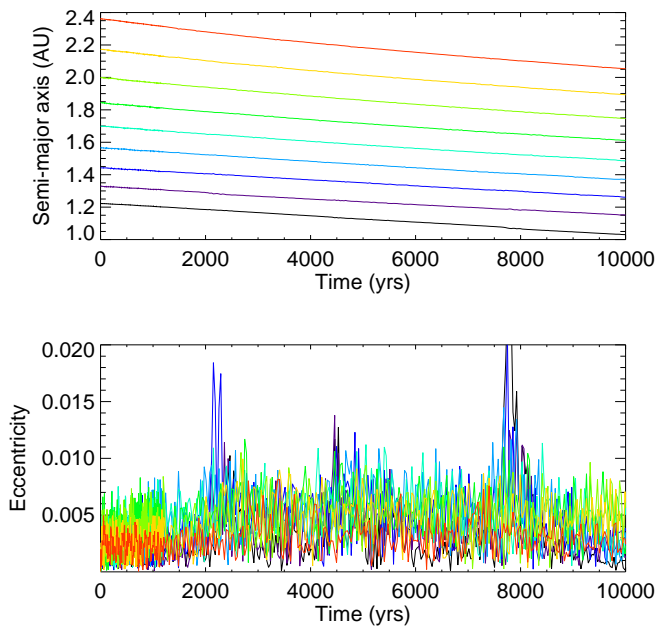}
\caption{Time evolution of the semimajor-axes and eccentricities of protoplanets for an isothermal hydrodynamical simulation 
with N=9.}
\label{fig:iso}
\end{figure}

In order to unveil the role of the zero-migration line on the collisions events that were observed in this simulation, 
we have performed a similar simulation with initially N=9 embryos but using an isothermal equation of state. In that case, all protoplanets are 
expected to experience inward migration due to their interaction with the gas disc. Fig. \ref{fig:iso} shows the evolution 
of the planets' semimajor-axes and eccentricities as a function of time for the isothermal run. We note that although we 
consider here equal-mass planets, these tend to undergo  convergent migration because the surface density profile has 
index $\sigma < 3/2$ (e.g. Pierens et al. 2011). Compared with the radiative simulation, however, the convergent 
migration rate is sufficiently weak in the isothermal case to prevent close encounters between embryos. This clearly demonstates 
that near the transition between the outward and inward migration regimes,  close encounters and collisions can be stimulated due to the strong differential migration experienced by the 
embryos there.

\subsection{Effects of disc turbulence}
\label{sec:hydro-turb}
\begin{figure}
\centering
\includegraphics[width=\columnwidth]{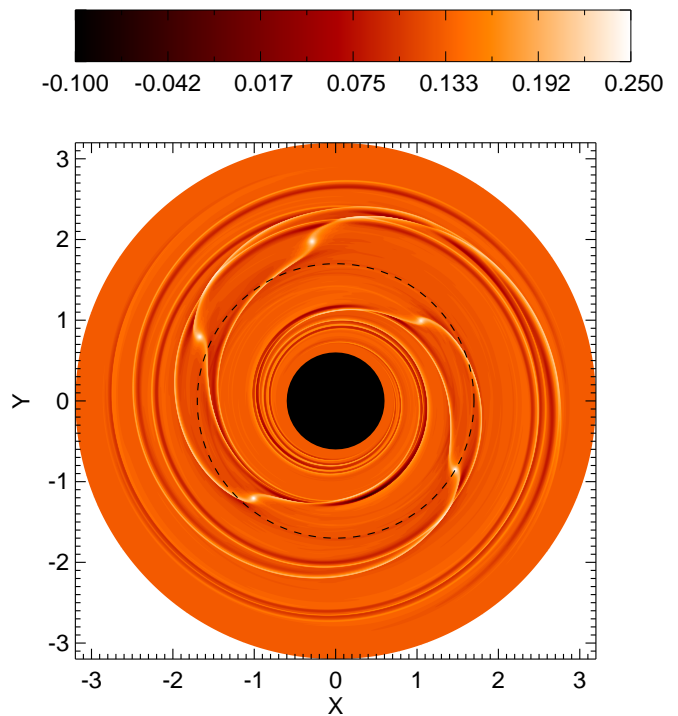}
\caption{Snapshot of the perturbed surface density profile in the hydrodynamical simulation with initially N=5 embryos and 
with stochastic forces included. The dashed circle shows the location of the zero-migration line.}
\label{fig:contour}
\end{figure}
\begin{figure}
\centering
\includegraphics[width=\columnwidth]{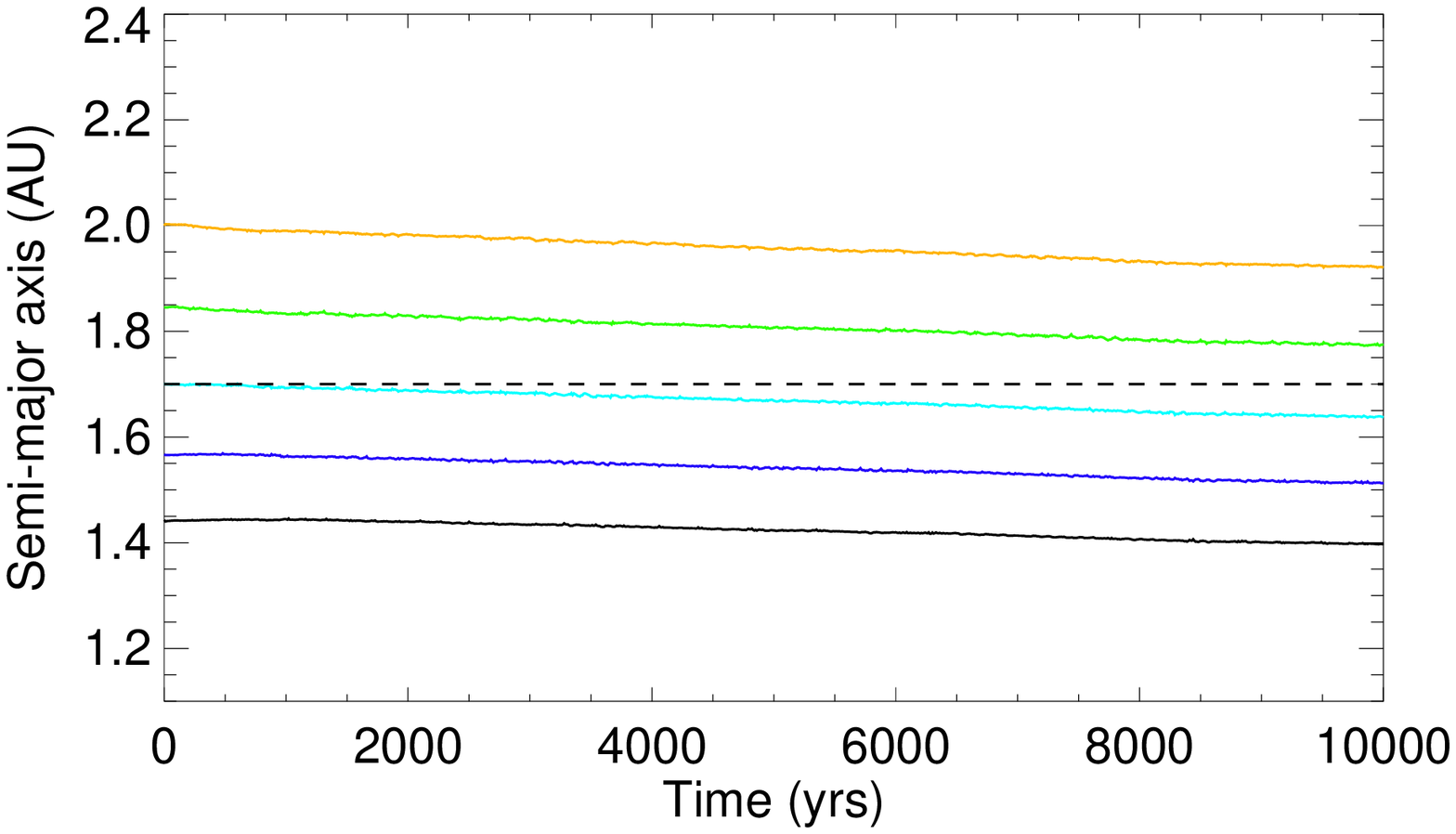}
\includegraphics[width=\columnwidth]{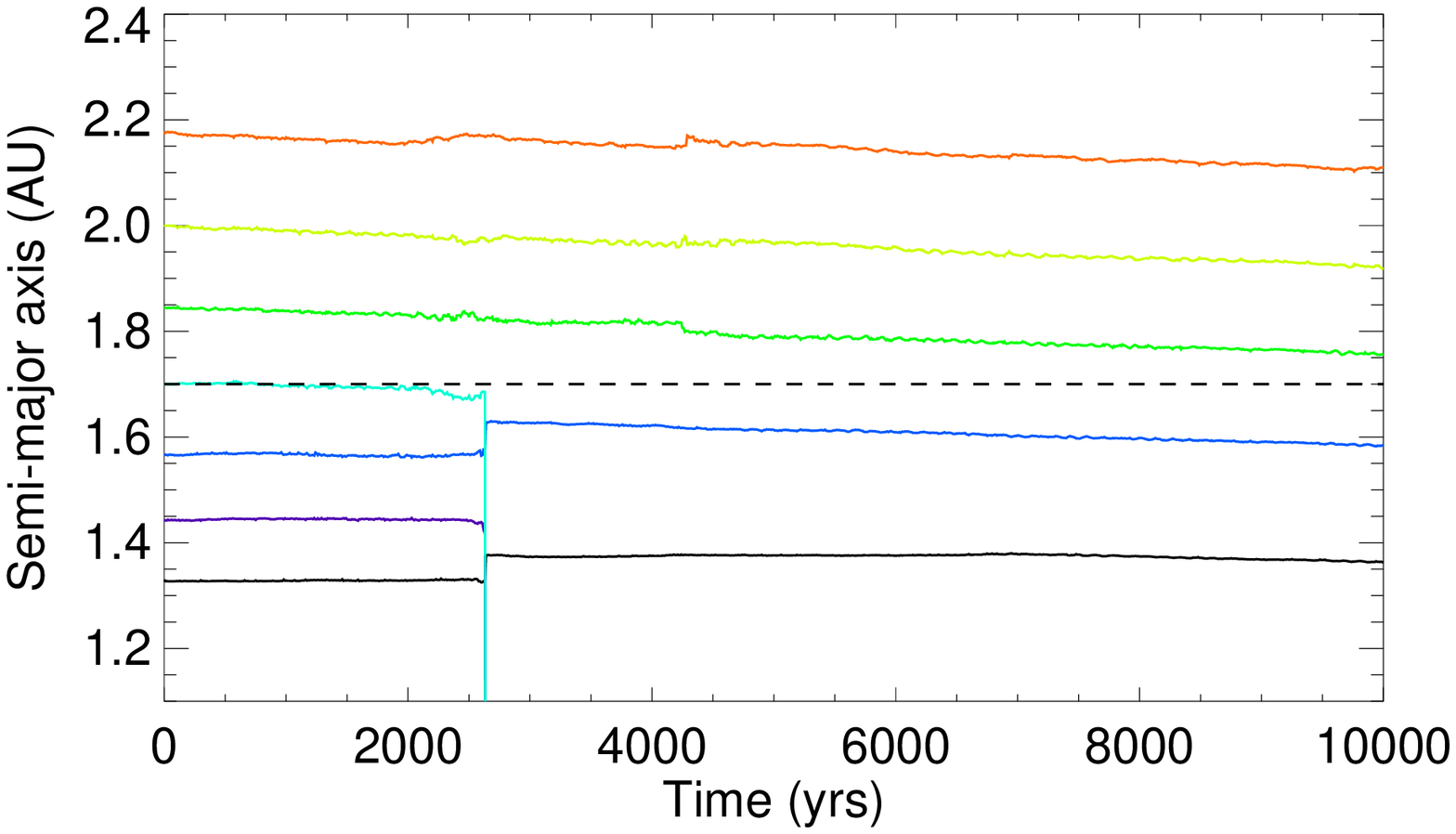}
\includegraphics[width=\columnwidth]{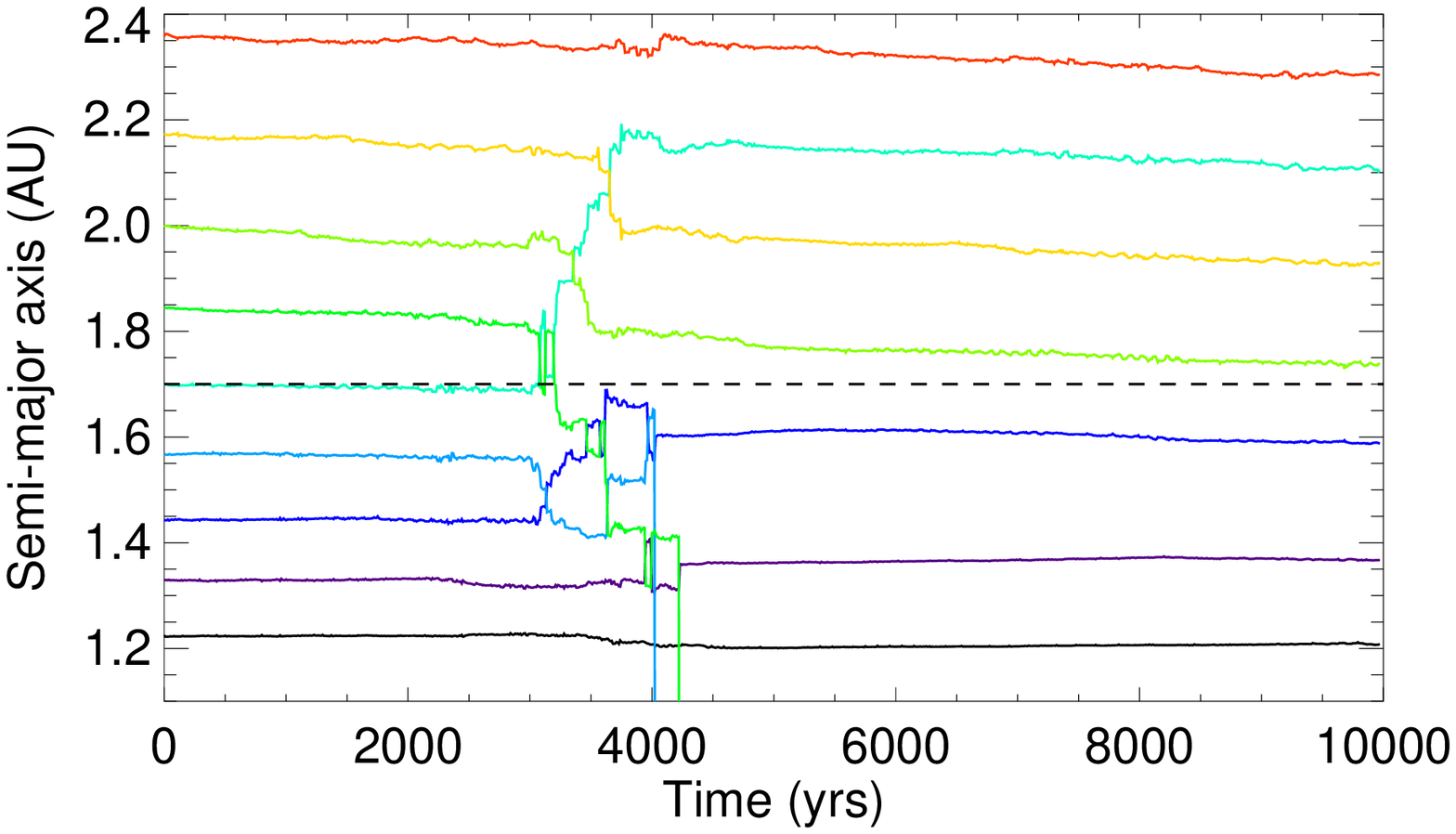}
\caption{{\it Top panel:}: time evolution of the semimajor-axes of protoplanets in a 
hydrodynamical simulation with N=5 and with 
stochastic forces included. {\it Middle panel:} same but for N=7. {\it Bottom panel:} same 
but for N=9.  The dashed line
 shows the location of the zero-migration line. }
\label{fig:turb}
\end{figure}
The results presented above indicate that, in the limit of a moderate initial number of objects, the formation of mean motion resonances prevents embryos from undergoing 
close encouters with other bodies. In this section, we examine how
the stability of these resonant configurations is affected  when a level of  disc turbulence is accounted for. 
Previous work (e.g. Pierens et al. 2011) suggested that mean motion resonances are likely to be disrupted 
by stochastic torques in the active regions of protoplanetary discs and within disc lifetimes. \\
To examine this issue, we have perfomed a series of simulations for N=5,7,9 in which each body is subject to a an 
additional stochastic force $\vec{F}_{\text{turb}} =-\nabla \Phi_{\text{turb}}$ where $\Phi_{\text{turb}}$ is given 
by Eq. \ref{eq:phi}. The temporal evolution of the planets' semi-major axis for these 
three simulations is displayed in Fig.\ref{fig:turb}. For N=5, Fig. \ref{fig:contour} presents 
a contour plot of the perturbed surface density distribution at the beginning of the simulation. 
For protoplanets initially located inside the zero-migration radius, surface density 
perturbations inside the planets' horseshoe regions and related to the co-orbital 
dynamics are clearly visible. Due to the negative entropy gradient, co-orbital 
dynamics leads to a positive (resp. negative) surface density pertubation ahead (resp. behind) 
of the planet, giving rise to a positive corotation torque. For protoplanets located outside 
the zero-migration line, these additional surface density perturbations are weaker,  
indicating a relatively faint corotation torque in that case. For this run, a sequence of 
9:8 MMRs is ultimately formed so that the final fate of the 
system is similar to that obtained in the laminar simulation. Although these resonances 
are stable on average, the corresponding resonant angles are observed to oscillate 
between periods of circulation and libration, implying a weaker resonant locking 
in presence of turbulence. We notice that a similar behaviour was  observed in previous studies on the capture in 
resonance of pairs of planets embedded in a turbulent isothermal disc (Pierens et al. 2011). This 
stengthens the expectation that turbulence can break resonant configurations when  the 
typical amplitude of the stochastic density fluctuations is large enough,  and consequently that the resonant systems 
 obtained using viscous laminar disc models are  more prone to destabilization in presence of 
turbulence. This is therefore not too surprising that collisions arise in the turbulent run with N=7, 
as can be seen in the middle panel of Fig. \ref{fig:turb}. Here, the resonant system that is formed at the end of the simulations 
is composed of two $6\;M_\oplus$ bodies (black+blue lines) located at the inner edge of the swarm and trapped in a 
5:4 resonance plus three exterior $3\; M_\oplus$ protoplanets trapped in a first-order p+1:p resonance with its 
neighbours, where we observe a clear tendency for $p$ to increase as  one moves out through the swarm.\\
The lower panel of Fig. \ref{fig:turb} shows the evolution of the system for the run with N=9. Again, two $6\;M_\oplus$ 
planets (purple+blue) are formed in the course of the evolution and the final fate corresponds again to the formation of a resonant chain with the 
two more massive planets located in the inner half of the swarm. Although three collisions 
ocurred in the laminar run with N=9, it is clear,  when comparing Fig. \ref{fig:turb} with Fig. \ref{7mp3}, that the evolution of the system is much more chaotic
 in the turbulent case. 
\subsection{Effect of the disc model}
\begin{figure}
\centering
\includegraphics[width=\columnwidth]{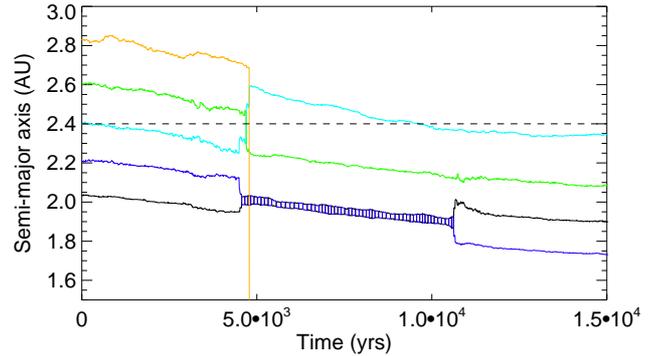}
\caption{Time evolution of the semimajor-axes of protoplanets in a 
turbulent hydrodynamical simulation with initially N=5 embryos and for a disc mass equivalent to two times 
that of the fiducial disc model.  The dashed line
 shows the location of the zero-migration line.}
\label{fig:model3}
\end{figure}
To investigate how the disc model affects the results presented above, we have performed a series of 
 simulations using a disc model with mass twice that of the fiducial disc model, which corresponds to 
a the snow-line  located at $\sim 4.2$ AU.  Again, the initial mass of embryos is 
$3\; M_\oplus$ and their initial positions are 
chosen such that half of the initial population of embryos 
is located inside the  zero-migration line, which lies at $\sim 2.4$ AU for this disc model, whereas 
the other half is located outside. Although not shown here, runs with stochastic forces not included show  evolution outcomes 
 consistent with those of the fiducial disc model, with  growth of embryos and  formation of $6\; M_\oplus$ protoplanets 
observed when  the initial number of bodies is $N \ge 7$. This is not surprising since as the disc 
mass increases, the effect of 
a faster differential migration toward the zero-migration line, and which would promote close encounters between embryos,
 is couterbalanced by a stronger disc-induced eccentricity damping. \\
For runs with stochastic forces included, however, the outcomes are found to differ noticeably from those of the fiducial 
disc model due to more vigorous turbulent fluctuations. In that case,  collisions are found to arise even for low values 
of the initial number of embryos, as illustrated in Fig. \ref{fig:model3}
where is displayed, for this disc model, the evolution of the planets' semimajor-axes for N=5. Here, a 
$6\;M_\oplus$ planet (cyan) is formed at the outer edge of the swarm at $t\sim 5\times 10^3$ yr and is likely 
to become trapped at the  location of the zero-torque radius at later times. From $t\sim 1.3\times 10^4$ yr, this planet tends to separate from the   
three innermost $3\;M_\oplus$ bodies which experience resonant inward migration with each planet forming a 8:7 resonance 
with its neighbours. Prior to the formation of this inner resonant system, it is interesting to note that the two 
innermost bodies (black+blue) entered in a coorbital 1:1 resonance and which remained stable for $\sim 5\times 10^3$ yr.\\
Again, the final fate of the system remains uncertain but it is likely that the group composed of the three innermost 
planets will end up on a stationary orbit once the net torque acting on this three-planet system will cancel out (Cossou et al. 2013) 
\section{Results of N-body runs}
\label{sec:nbody}
In Sect. \ref{sec:calibration}, we have presented how  analytical formulae for the Type I torques 
can be calibrated using the  results of hydrodynamical simulations.  Here, we 
present the main results that emerge from  $\sim 600$ N-body runs of protoplanets 
embedded in a radiative protoplanetary disc that we have performed using the prescriptions we 
obtained for Type I migration. The aims of this alternative approach are to i) study the long-term 
evolution of a swarm of protoplanets migrating toward the convergence line and ii) examine 
the statistical properties of the planetary systems which are formed through this process.  
\subsection{Comparison with hydrodynamical simulations }
\subsubsection{Runs without stochastic forces included}
We first describe the range of outcomes that are observed in simulations with initial conditions 
chosen as close as possible to those for the hydrodynamical simulations presented in 
Sect. \ref{sec:hydro}. 
We performed 100 N-body simulations for situations with N=7 and N=9 initial embryos. These simulations  
did not include stochastic forcing and differed only in the initial azimuthal positions 
of the embryos. For N=7, collisions occured in 53\% of the simulations. Objects as
large as 9 $M_\oplus$ formed in 6\% of the simulations. In the runs with N=9, virtually 
all (95\%) of the simulations included collisions, as expected from the hydrodynamical 
simulations presented in Sect. \ref{sec:hydro}. Most (77\%) simulations only produced 
6 $M_\oplus$ planets, but a significant number (17\%) formed 9 $M_\oplus$ planets and 
one simulation formed a 12 $M_\oplus$ core.

\begin{figure*}
\centering
\includegraphics[width=\textwidth]{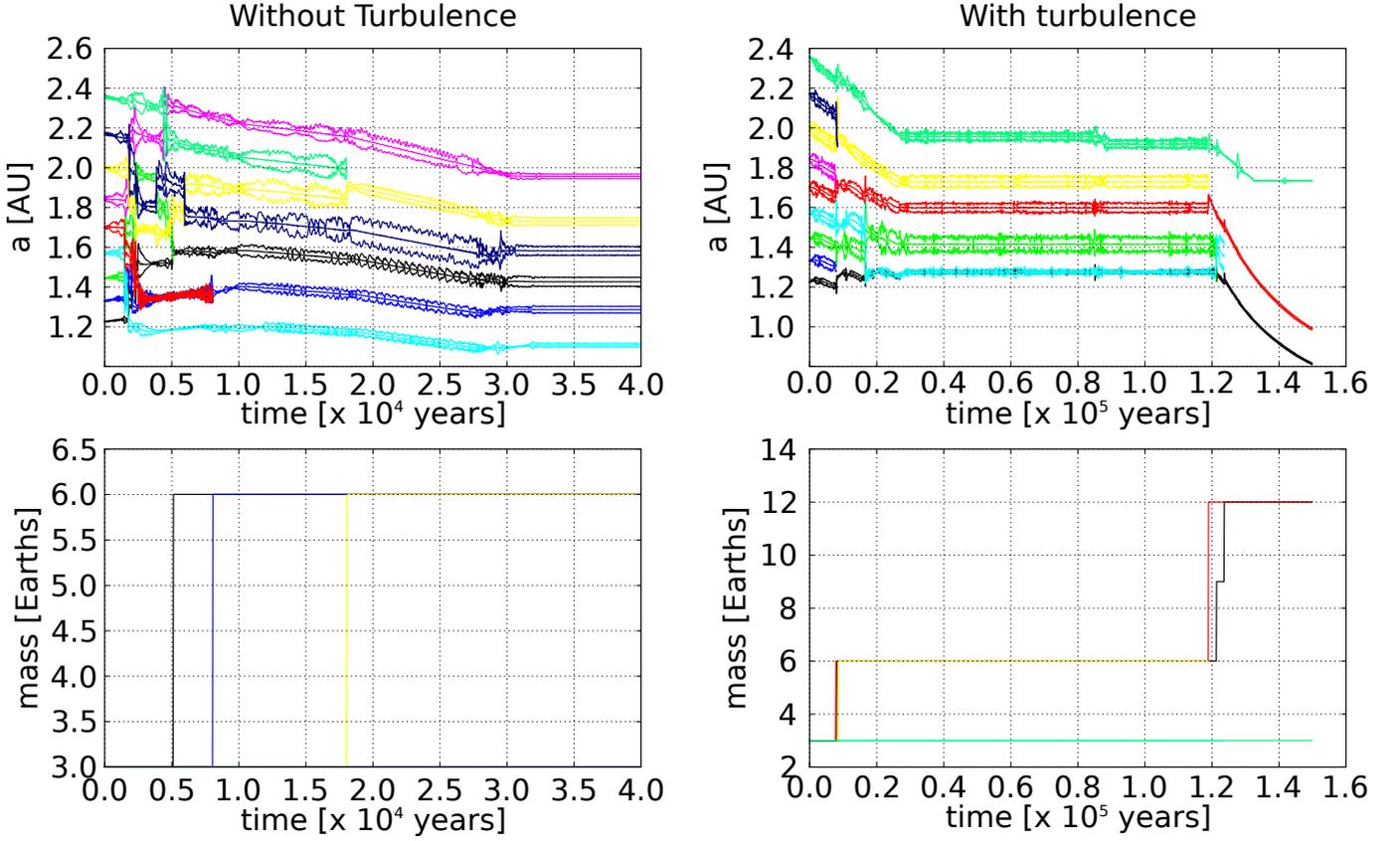}
\caption{{\it Left:} time evolution of protoplanets' orbital positions (top panel) and 
masses (lower panel) for a 
N-body run with intially N=9 embryos and without stochastic forces included. 
{\it Right:} same but with stochastic forces included. }
\label{fig:nbody9}
\end{figure*}

The left panel of Fig. \ref{fig:nbody9} displays the evolution as a function of time of the 
 planets' orbital positions and masses for a run representative of the range of the observed outcomes. Overall, the early stages of evolution are 
consistent with what is seen in the corresponding hydrodynamical simulation, 
involving outward (resp. inward) migration of the innermost (resp. outermost) bodies plus 
formation of two $6\;M_\oplus$ protoplanets inside the zero-migration line by $\sim 10^4$ years.  
It is interesting to notice that of these two $6\;M_\oplus$, one is formed 
at $\sim 8\times 10^3$ years  
through  the merging of two co-orbital planets that entered
in a 1:1 resonance   at $\sim 2\times 10^3$ years. An additional $6\;M_\oplus$ body is produced 
at $t\sim 1.8\times 10^4$ yr as a result of the collision between two $3\;M_\oplus$ objects. After $\sim 3\times 10^4$ years, the systems attains 
a quasi-stationary state and consists of three $6\;M_\oplus$ objects plus 
two $3\;M_\oplus$ bodies trapped in a resonant chain 
and evolving on non-migrating orbits. Moving from inward to outward, the 
resonances which are formed are 5:4, 7:6, 7:6, 8:7 and 6:5. This stationary configuration is reached when the 
positive torque exerted on the two innermost $6\;M_\oplus$ bodies (blue and black 
lines in Fig. \ref{fig:nbody9}) counterbalances the 
torques experienced by the three other bodies, leading to a zero net torque 
acting on the whole system (Cossou et al. 2013). 
\subsubsection{Effect of stochastic forces}
\label{sec:nbody-turb}
To illustrate the dependence of the results presented above on the presence of disc turbulence, we have performed 
 a set of simulations but  with stochastic forces acting on the protoplanets included. We remind the reader that 
hydrodynamical simulations in which stochastic forces are included (see Sect. \ref{sec:hydro-turb})  indicated that collisions are more likely to occur 
in that case due to the general tendency for turbulent fluctuations to break resonant configurations. In 
agreement with this result, we find that collisions occured in $97\;\%$ of the N-body runs performed with N=7 and in 
$100\%$ of the simulations performed with N=9. Therefore, it is not surprising to observe 
a clear trend for forming more massive embryos when stochastic forces are included. For example, $36\;\%$ of the runs with 
N=7 resulted in the production of $9\;M_\oplus$ bodies whereas this number increases to $50\;\%$ for 
N=9. In this latter case,  two simulations resulted in the formation of $12\;M_\oplus$ protoplanets. Moreover, 
 we note that we performed a series of turbulent runs 
with initial separations of $6\;R_{mH}$, and which resulted in the production of $9\;M_\oplus$ cores in $30$ of a
 total of  $100$ 
simulations, which suggests that these statistics are relatively robust with regards to  the value for the initial 
separation of embryos.\\
The right panel of Fig. \ref{fig:nbody9} illustrates the evolutionary outcome for a 
simulation in which two $12\;M_\oplus$ planets are produced. At 
$t\sim 10^5$ yr, a non-migrating compact system is formed with two co-orbital planets 
of $3\;M_\oplus$ and $6\;M_\oplus$ located at the inner edge of the swarm and in a 
7:6 resonance with an exterior $3\;M_\oplus$ protoplanet. At 
$t\sim 1.2\times 10^5$ years, the formation of a $12\;M_\oplus$ body through the 
collision of two exterior $6\;M_\oplus$ bodies (yellow+red) 
destabilizes the inner part of the swarm, which subsequently leads to the 
merging of the three inner planets into an additional $12\;M_\oplus$ embryo. From this time, the system is composed of two $12\;M_\oplus$ bodies plus an exterior 
$3\;M_\oplus$ embryo which evolves on a quasi-circular orbit at the nominal 
convergence line. The two $12\:M_\oplus$ planets enter in a 6:5 resonance at later times and resonantly migrate 
inward until they reach the inner edge of the disc. This configuration remained stable until the end 
of the simulation which was evolved for $5\times 10^5$ yr.    We note that although these planets evolve inside 
the zero-migration line, they experience a negative torque from the disc because 
the (positive) corotation torque is saturated for such a planet mass (see Sect. \ref{sec:calibration}).
\subsection{Effect of changing the initial mass distribution and statistical overview}
\begin{figure*}
\centering
\includegraphics[width=0.49\textwidth]{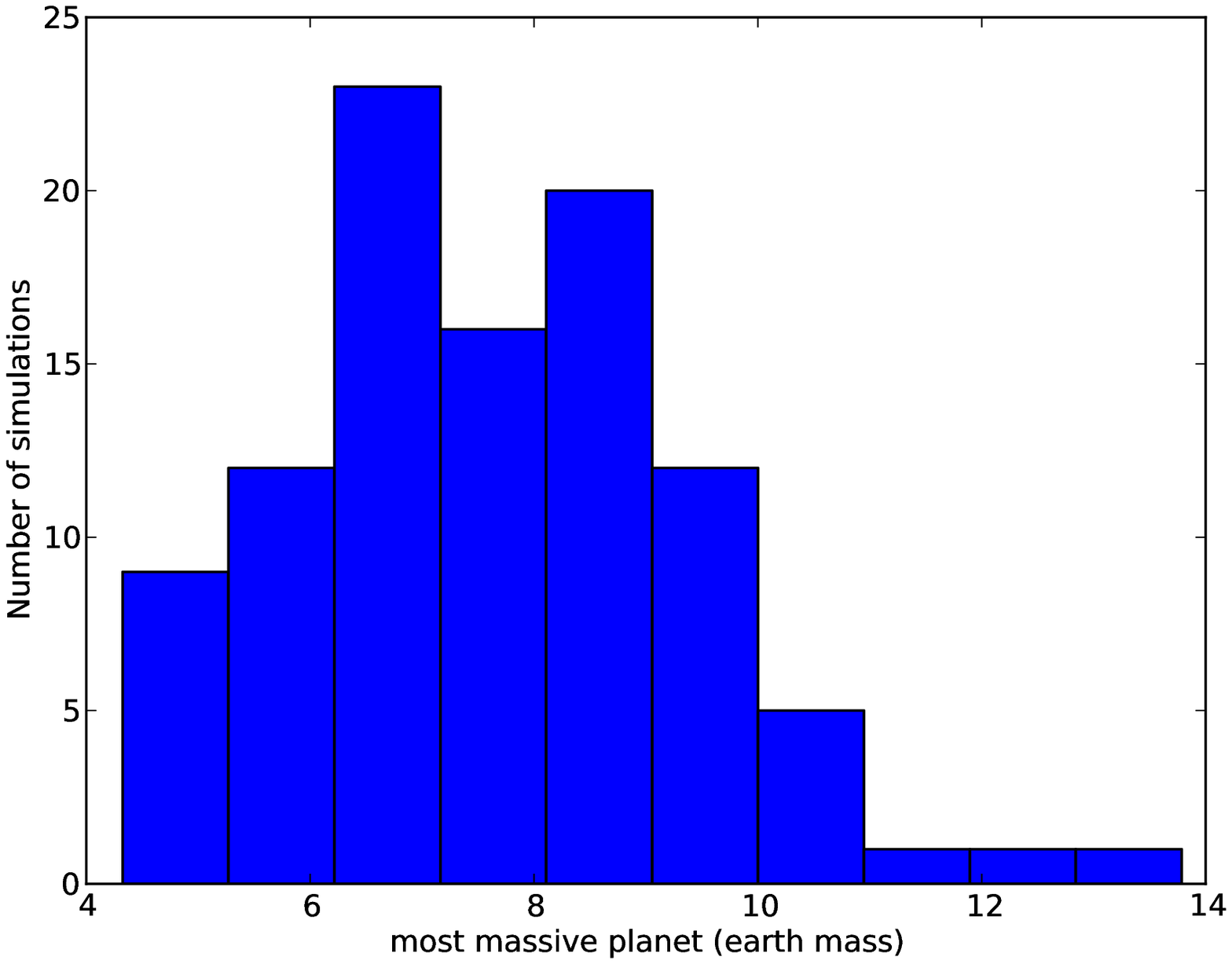}
\includegraphics[width=0.49\textwidth]{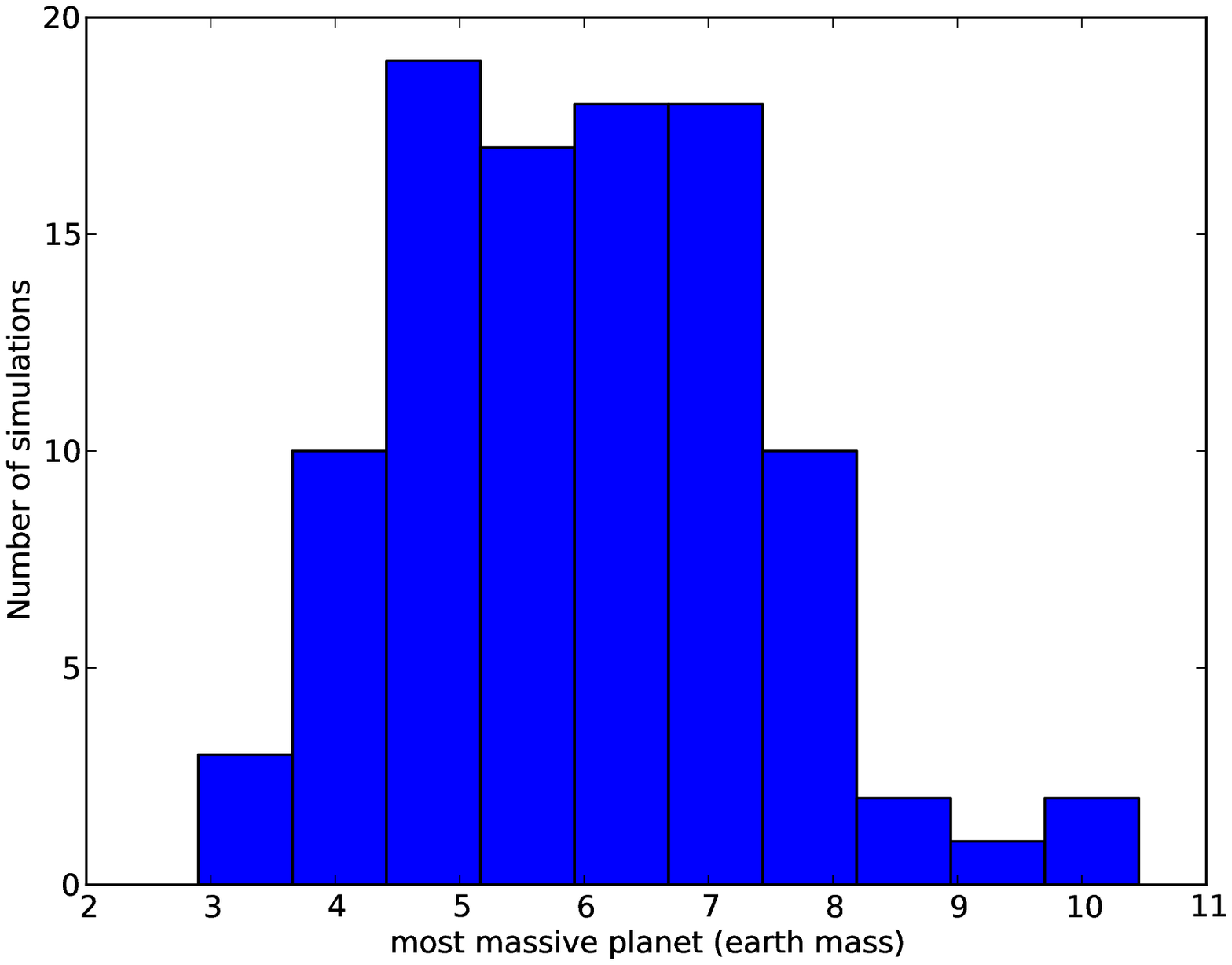}
\caption{{\it Left panel:} distribution of mass of the most massive planet which is formed in N-body runs with stochastic 
forces included and in which  
the mass of each embryo is sampled from a Gaussian distribution with $\mu_m=3\;M_\oplus$ and $\sigma_m=1\;M_\oplus$. 
{\it Right panel:} same but for $\mu_m=1\;M_\oplus$ and $\sigma_m=0.5\;M_\oplus$.}
\label{fig:histo_mass}
\end{figure*}

In order to examine the dependence of our results on the initial mass distribution, we performed two 
additional sets of $100$ simulations using a randomised mass distribution. In the first 
series of runs, the 
mass of each embryo is sampled from a Gaussian distribution with mean $\mu_m=3\;M_\oplus$ and 
standard deviation $\sigma_m=1\;M_\oplus$, while in the second set of simulations, we 
set $\mu_m=1\;M_\oplus$ and $\sigma_m=0.5\;M_\oplus$. The total mass of embryos is chosen 
to be $30\;M_\oplus$ in both cases. Moreover,  we choose the orbital position of the inner planet to 
be uniformely distributed between $1.1\le a_p \le 1.2$ AU whereas the initial separations of 
 other planets are set to $nR_{mH}$, where $4\le n \le 5$ is randomly chosen according to 
a uniform distribution.\\
Fig. \ref{fig:histo_mass} shows the distribution of mass of the most massive planet which is formed in 
these two series of simulations. In the case where $\mu_m=3\;M_\oplus$ and 
$\sigma_m=1\;M_\oplus$ (left panel), we see that $\sim 7\;M_\oplus$ embryos are most common but of the $100$ simulations, 
a significant number of them ($43\%$)  
 resulted in the formation of protoplanets with masses in the range $8 \le m_p\le 10\;M_\oplus $ within 
$5\times 10^5$ yr. This is  
in good agreement with the results of simulations with initially equal-mass embryos presented in 
Sect. \ref{sec:nbody-turb}. Giant planet cores 
with masses $\ge 10\;M_\oplus$ were produced in $ 8\;\%$ of the simulations.\\
In the case where $\mu_m=1\;M_\oplus$ and $\sigma_m=0.5\;M_\oplus$ (right panel),  $14\;\%$ of runs resulted 
in the formation of bodies with masses in the range $8\le m_p\le 10\;M_\oplus$, with only one 
run leading to the formation of a  core with mass $\ge 10\;M_\oplus$.  We believe this is related to the fact 
that the initial population of embryos with initial masses $\lesssim 1\;M_\oplus$ 
experience corotation torque saturation (see Sect. \ref{sec:calibration}). Consequently, 
they undergo inward Type I migration even these 
are located inside the opacity transition, which tends to substantially reduce the process of convergent 
migration at the planet trap.\\
These results indicate that  forming giant planet cores at the zero-torque radius is likely to occur provided it involves massive impacts between bodies of a few Earth masses which do not 
experience corotation torque saturation and which formed earlier through an alternative process.
\section{Discussion}

\begin{figure}
\centering
\includegraphics[width=\columnwidth]{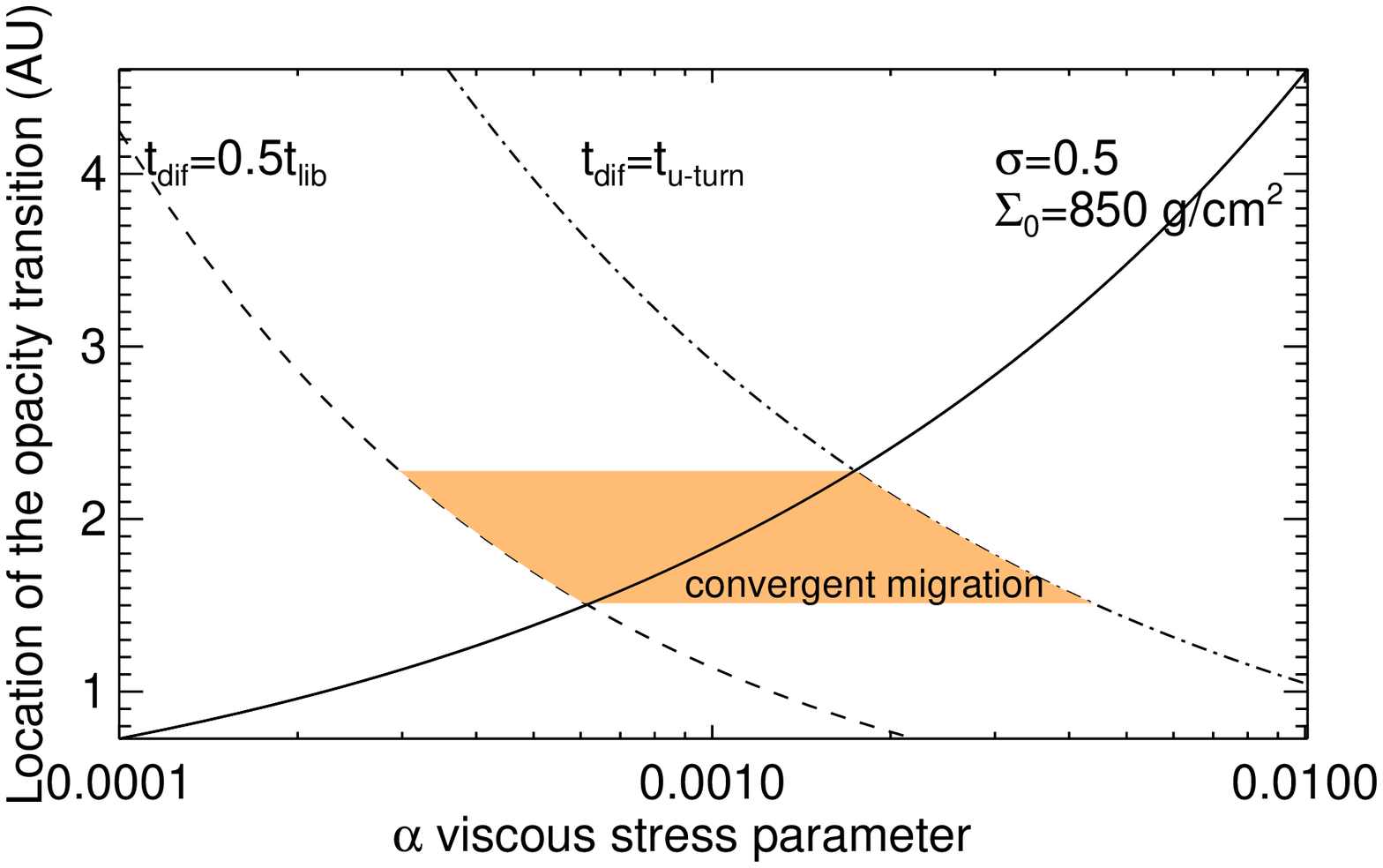}
\includegraphics[width=\columnwidth]{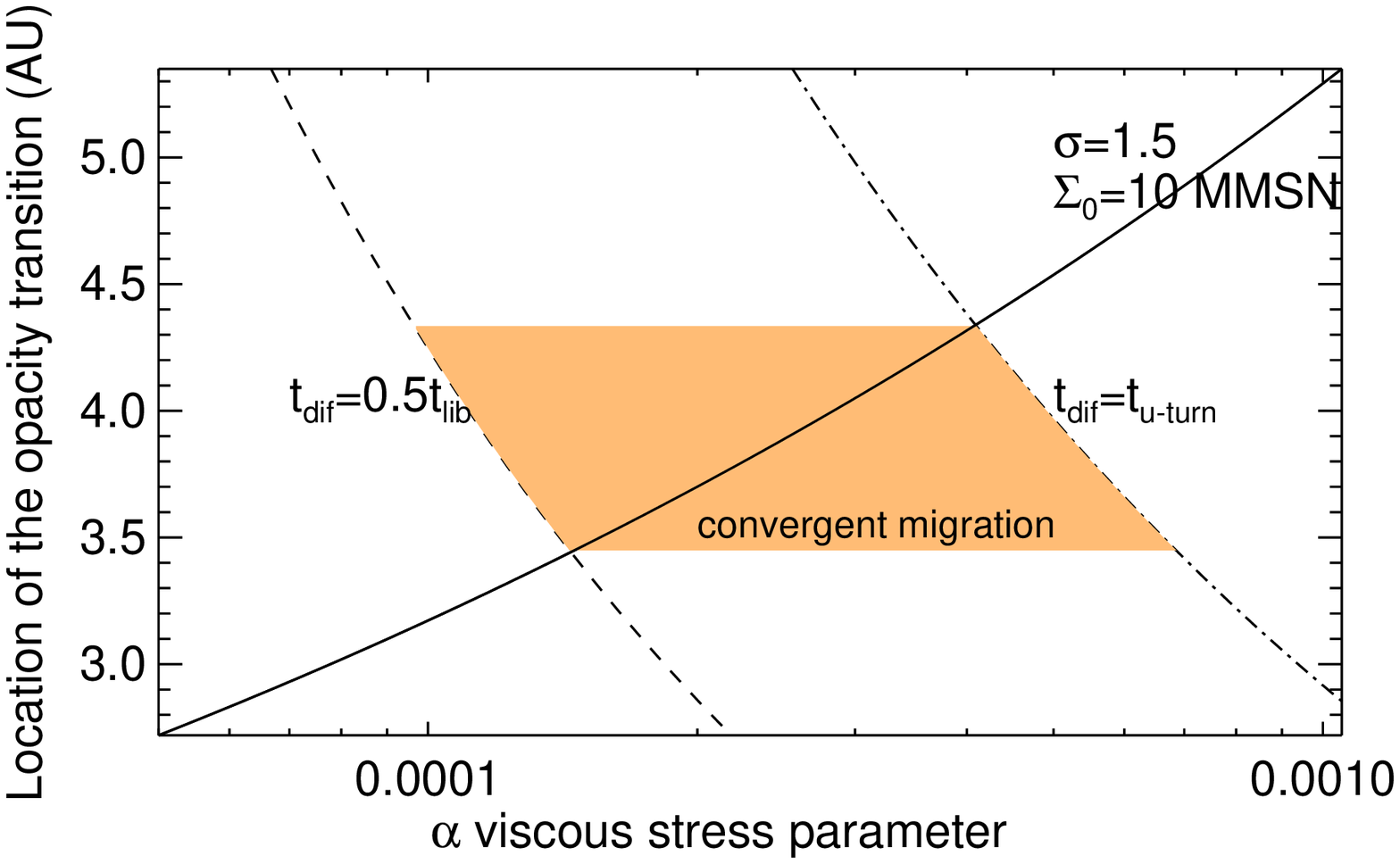}
\caption{{\it Top panel:} Location of the opacity transition (solid line) as a function of the value for the $\alpha$ 
 viscous stress parameter and for our fiducial disc model. The dashed line shows the location where the viscous 
timescale $t_{dif}$ becomes equal to half of the horseshoe libration timescale ($t_{lib}$) whereas the solid-dashed line 
shows the radius where $t_{dif}$ equals the U-turn timescale. The intersection between the shaded area and the 
solid line corresponds to the range of radii where 
convergent migration for $3\;M_\oplus$ bodies is expected to occur. {\it Lower panel:} same but for a disc model 
ten times more massive than the MMSN.}
\label{fig:discussion}
\end{figure}

In this section,  we discuss how the location 
of the opacity transition we considered here depends on the disc model that is adopted and examine under which conditions 
collisional planetary growth at this opacity transition is expected to arise.\\
In order to determine the location of the opacity transition, we balance radiative losses with viscous 
heating, which gives:
\begin{equation}
2\sigma T_{\text{eff}}^4=\frac{9}{4}\nu\Sigma\Omega^2
\label{eq:base}
\end{equation}
For an optically thick disc, we remind that the midplane temperature $T$ is related to the effective temperature 
$T_{eff}$ by (see Eq. \ref{eq:tau}):
\begin{equation}
T^4=\frac{3}{8}\tau T_{\text{eff}}^4
\end{equation}
Given that $\tau=\kappa\Sigma/2$, Eq. \ref{eq:base} can be recast as:
\begin{equation}
\sigma T^4=\frac{27}{128}\nu\kappa\Sigma^2\Omega^2
\end{equation}
Moreover, in the context of the standard $\alpha$ prescription of Shakura \& Sunyaev (1973) , the 
effective kinematic viscosity is given by $\nu=\alpha c_s H$, where $c_s=(\gamma_{ad} {\cal R} T/\mu)^{1/2}$ 
(${\cal R}$ and $\mu$ are the gas constant and the mean molecular weight respectively) is the 
sound speed and $H=c_s/\Omega$ the disc scale height. Evaluating the previous equation
at the opacity transition gives the following expression for the location of the transition 
radius $R_t$:
\begin{equation}
\frac{R_t}{R_0}=\left(\frac{27}{128}\frac{\gamma_{ad} {\cal R}}{\mu \sigma}\frac{\kappa_t\Sigma_0^2\Omega_0}{T_t^3}\alpha \right)^{1/(2\sigma+3/2)}
\label{eq:loc}
\end{equation}
where  $T_t$ and $\kappa_t$ denote the values for the
temperature and opacity at the transition respectively. Here, we have used the fact that $\Sigma=\Sigma_0(R/R_0)^{-\sigma}$ and $\Omega=\Omega_0(R/R_0)^{-3/2}$, where 
$\Omega_0$ is the angular velocity at $R=R_0$. \\
Furthermore, convergent migration at the opacity transition for bodies of a few Earth masses  arises provided that the 
corotation torque remains unsaturated. We expect the corotation torque to remain close to its 
fully unsaturated value provided that (e.g. Baruteau \& Masset 2013):
\begin{equation}
t_{\text{U-turn}}\le t_{\text{dif}}\le t_{\text{lib}}/2
\end{equation}
For given values of $\alpha$ and the mass ratio $q=m_p/M_\star$, this condition provides an estimation of the range of radii for which 
the corotation torque is fully unsaturated. We find:
\begin{equation}
0.52\left(\frac{\Omega_0R_0}{c_{s,t}}\right)^2q^{6/7}\alpha^{-4/7}\le\frac{R}{R_0}\le0.025\left(\frac{\Omega_0R_0}{c_{s,t}}\right)^2q^{2/3}\alpha^{-4/9} 
\label{eq:satur}
\end{equation}
where $c_{s,t}$ is the value of the sound speed at the opacity transition. In the top panel of Fig. \ref{fig:discussion} we plot the location of $R_t$ as a function of the $\alpha$ 
viscous stress parameter and for the disc model that we employed in the simulations. The region located in between the dashed 
and solid-dashed lines in Fig. \ref{fig:discussion} corresponds to that  where the corotation torque 
remains unsaturated for a $3\;M_\oplus$ planet, and which is defined by Eq. \ref{eq:satur}. Therefore, 
the intersection between the solid line and the shaded area in Fig. \ref{fig:discussion} represents the range 
of radii where convergent migration of $3\;M_\oplus$ protoplanets  can occur. 
For this disc model, this process is expected to arise in the region from $\sim 1.5$ to $\sim 2.2$ AU, in good agreement with the 
location of the zero-torque radius in our hydrodynamical and N-body simulations.\\
 For a given value of $\alpha$, Eq. \ref{eq:loc} predicts that  the location of the opacity transition 
will  increase with  $\Sigma$. However, looking at the upper panel of Fig. \ref{fig:discussion}, it is clear 
that as $\Sigma$ increases, the range of $\alpha$ values for which the corotation torque remains unsaturated 
is shifted toward lower values. In the context of the formation of the giant planets in the 
Solar System, this implies that both a significantly massive disc and a  relatively low value for $\alpha$
  will be needed  for the convergent migration mechanism to operate in the Jupiter-Saturn region. 
This is illustrated in the lower panel of Fig. \ref{fig:discussion} which shows that for a disc model 
with $\sigma=1.5$, convergent migration at $\sim 4$ AU arises provided that the disc mass corresponds to ten times the MMSN. 
Of course, these results will strongly depend on the opacity transition that is considered. We nevertheless expect a 
similar mechanism to occur at any opacity transition interior to which the temperature gradient is steep enough 
to allow for outward migration of Earth-mass bodies. Testing other opacity transitions is beyond the scope of that 
paper but we will discuss in a future paper the influence of changing the opacity table. Finally, we  notice that  formation of giant planet cores  through
collisions of Earth-mass bodies may also be possible  at other kinds of planet traps like those arising at a dead zone or at the radius where stellar heating begins to take over viscous heating (Hasegawa \& Pudritz 2011).

\section{Conclusion}
We have presented the results of both  hydrodynamical and N-body simulations of the 
evolution of a swarm of Earth-mass protoplanets that gravitationally interact near a  
zero-migration line. 
The embryos are initially located around an opacity transition which plays the role 
of a planet trap where the Type I torque cancels. For bodies with masses 
in the range $1-10\;M_\oplus$ and evolving 
inside the opacity transition,  Type I migration proceeds outward whereas 
it is directed inward if these are 
located further out.  The main aim of this work is to examine the possible outcomes 
that arise when low-mass protoplanets convergently migrate toward a 
zero-migration radius and in particular  whether giant planet cores can be formed at such 
 places through giant impacts between embryos of a few Earth masses.\\ 
Hydrodynamical simulations show that  
equal-mass embryos with mass of $3\;M_\oplus$ located on both sides of a convergence zone tend to enter in a 
resonant chain whose stability depends on the initial number of objects 
and whether or not planets experience stochastic forces due to turbulence. For a limited 
number of bodies and in the absence of stochastic forcing, a sequence of resonances appears 
to be stable 
 so that close encounters betweeen embryos are prevented. Increasing the initial number 
of protoplanets however leads to a significant compression of the system and eventually to the 
destabilization of these resonant chains. Formation 
of $6\;M_\oplus$ protoplanets from $3\;M_\oplus$ embryos  is observed to occur in  
that case on a timescale $\le 10^4$ yr.\\
Not surprisingly,  including a moderate level of turbulence corresponding 
to a value for the $\alpha$ viscous stress parameter of $\alpha \sim 2\times 10^{-3}$ clearly enhances 
this process of collisional planetary growth. Interestingly, we find that a
 significant fraction ($\sim 50\;\%$) of the  N-body runs with stochastic forces included and performed 
with masses randomly sampled from a Gaussian distribution with $\mu_m=3\;M_\oplus$
resulted in the formation of giant planet cores with mass $\ge 8-10\;M_\oplus$ in $\sim 5\times 10^5$ yr. 
For a randomised mass distribution with $\mu_m=1\;M_\oplus$, however, only $\sim 15\;\%$ of the N-body 
runs produced giant planet cores. We conclude that forming giant planet cores at convergence zones is 
efficient provided that it involves collisions between embryos with mass $\gtrsim 2\;M_\oplus$ and
which formed earlier according to the classical 
runaway/oligarchic growth scenario (Levison et al. 2010) or through accretion of cm-sized pebbles 
(Morbidelli \& Nesvorny 2012). \\
For a protoplanetary disc with mass typical of the MMSN, we find that the mechanism presented here may allow the {\it in-situ} formation of giant planet cores from Earth-mass bodies at $\sim 2$ AU  whereas very massive discs ($\ge 10$ times the MMSN) 
are required to form $\sim 10\;M_\oplus$ bodies at $4-5$ AU. We note however that we considered here  
a zero-migration line corresponding to  a particular change in the opacity regime. Other planet traps 
located further than $\sim 4$ AU are expected to arise in typical protoplanetary disc models. For example, the transition 
where stellar irradiation begins to provide most of the heating of the disc gives rise to an additional planet trap located at 
$2-40$ AU, depending on the mass accretion rate (Hasegawa \& Pudritz 2011). Planet traps can also exist at locations  
where the thermal diffusion timescale becomes longer than the horseshoe libration timescale, leading to a saturation of the 
corotation torque in the outer regions. Since the horseshoe libration timescale depends on the planet 
mass, planets with different masses tend to converge toward different radii. We will focus on the role of these 
additional planet traps on the formation of giant planet cores in a future publication.\\
The N-body runs that we have presented are the simplest we can perform.   One limitation of our work 
is that the embryos do not accrete planetesimals or pebbles as they migrate.  If there is a sufficient supply of pebbles, 
accretion may be very rapid for embryos of $\sim 1\;M_\oplus$ (Lambrechts \& Johansen 2012; Morbidelli \& Nesvorny 2012), with the consequence that embryo masses may  change significantly over the timescales of the 
simulations. Morevover, the position of the zero-migration line remains fixed in the course of the simulations. In a more 
realistic scenario, we expect the convergence zone to move inward as the disc disperses (Lyra et al 2010; Horn et al 2012). We will present in a forthcoming paper the results of both hydrodynamical and N-body simulations that account for  
the inward migration of the zero-torque radius due to  photoevaporation and irradiation from the central star effects.

\begin{acknowledgements}
Computer time for this study was provided by the computing facilities MCIA (M\'esocentre de Calcul Intensif Aquitain) of the Universit\'e de Bordeaux and by HPC resources from GENCI-cines (c2012046957).

\end{acknowledgements}

\label{lastpage}

\end{document}